\documentclass{article}

\PassOptionsToPackage{numbers, compress}{natbib}

\usepackage[preprint]{neurips_2025}
\usepackage{longtable}

\usepackage[utf8]{inputenc} 
\usepackage[T1]{fontenc}    
\usepackage{hyperref}       
\usepackage{url}            
\usepackage{enumitem}      
\usepackage{booktabs}       
\usepackage{amsfonts}       
\usepackage{amsmath}
\usepackage{nicefrac}       
\usepackage{adjustbox} 
\usepackage{microtype}      
\usepackage{xcolor}         
\usepackage[capitalise]{cleveref}
\usepackage{algorithm}
\usepackage{algpseudocode}
\usepackage{pgfplots}
\usepackage{tikz}
\usetikzlibrary{shapes, arrows.meta, positioning}

\def\removecomments{0} 
\if \removecomments1

\fi

\usepackage{subcaption}
\pgfplotsset{width=5.5cm,compat=1.9}
\usepackage{multirow}
\usepackage{makecell}
\usepackage{color, colortbl}
\usepackage[flushleft]{threeparttable}
\usepackage{adjustbox}
\usepackage{float}
\definecolor{Gray}{gray}{0.9}
\definecolor{Blue}{rgb}{0.84,0.92,0.94}
\usepackage{tabularx}
\usepackage{changepage}
\usepackage{listings}

\usepackage{listings}
\usepackage{xcolor}
\definecolor{codegreen}{rgb}{0,0.6,0}
\definecolor{codegray}{rgb}{0.5,0.5,0.5}
\definecolor{codepurple}{rgb}{0.58,0,0.82}
\definecolor{backcolour}{rgb}{0.95,0.95,0.92}

\lstdefinestyle{mystyle}{
  backgroundcolor=\color{backcolour}, commentstyle=\color{codegreen},
  keywordstyle=\color{magenta},
  numberstyle=\tiny\color{codegray},
  stringstyle=\color{codepurple},
  basicstyle=\ttfamily\footnotesize,
  breakatwhitespace=false,         
  breaklines=true,                 
  captionpos=b,                    
  keepspaces=true,                 
  numbers=left,                    
  numbersep=5pt,                  
  showspaces=false,                
  showstringspaces=false,
  showtabs=false,                  
  tabsize=2
}

\DeclareUnicodeCharacter{2212}{-}
\usepackage{booktabs}

\definecolor{mycolor}{rgb}{0.122, 0.435, 0.698}
\makeatletter
\newcommand{\mybox}[1]{%
  \setbox0=\hbox{#1}%
  \setlength{\@tempdima}{\dimexpr\wd0+13pt}%
  \begin{tcolorbox}[colframe=mycolor,boxrule=0.5pt,arc=4pt,
      left=6pt,right=6pt,top=6pt,bottom=6pt,boxsep=0pt,width=\linewidth-4pt,center]
    #1
  \end{tcolorbox}
}

\title{Analyzing Latent Concepts in Code Language Models}

\author{%
  Arushi Sharma \\
  Iowa State University \\
  USA \\
  \texttt{arushi17@iastate.edu}
  \And
  Vedant Pungliya \\
  Iowa State University \\
  USA \\
  \texttt{vedant29@iastate.edu}
  \And
  Christopher J.~Quinn \\
  Iowa State University \\
  USA \\
  \texttt{cjquinn@iastate.edu}
  \And
  Ali Jannesari \\
  Iowa State University \\
  USA \\
  \texttt{jannesar@iastate.edu}
}

\begin{document}

\maketitle

\begin{abstract}
Interpreting the internal behavior of large language models trained on code remains a critical challenge, particularly for applications demanding trust, transparency, and semantic robustness. We propose Code Concept Analysis (CoCoA): a global post-hoc interpretability framework that uncovers emergent lexical, syntactic, and semantic structures in a code language model’s representation space by clustering contextualized token embeddings into human-interpretable concept groups. We propose a hybrid annotation pipeline that combines static analysis tool-based syntactic alignment with prompt-engineered large language models (LLMs), enabling scalable labeling of latent concepts across abstraction levels. We analyse the distribution of concepts across layers and across three finetuning tasks.  Emergent concept clusters can help identify unexpected latent interactions and be used to identify trends and biases within the model's learned representations. We further integrate LCA with local attribution methods to produce concept-grounded explanations, improving the coherence and interpretability of token-level saliency. Empirical evaluations across multiple models and tasks show that LCA discovers concepts that remain stable under semantic-preserving perturbations (average Cluster Sensitivity Index, CSI = 0.288) and evolve predictably with fine-tuning. In a user study on the programming-language classification task, concept-augmented explanations disambiguated token roles and improved human-centric explainability by 37 percentage points compared with token-level attributions using Integrated Gradients.
\end{abstract}

\section{Introduction}
\label{Introduction}

Neural models have significantly advanced the state of the art across software engineering (SE) tasks, including code understanding, generation, and translation. Despite their empirical success, these models remain largely opaque. Their black-box nature hinders interpretability, limiting trust and adoption—particularly in high-stakes domains. Existing evaluation metrics, such as CodeBLEU~\cite{ren2020codebleu}, primarily assess surface-level correctness or functional equivalence, offering limited insight into the internal abstractions that models rely on. To build robust and trustworthy systems, we must move beyond accuracy and what focus on the model's organization of relevant concepts. 

Interpretability methods in machine learning are typically categorized as either \textit{local} or \textit{global}. Local methods—such as SHAP~\cite{NIPS2017_7062}, LIME~\cite{lime}, and Integrated Gradients~\cite{sundararajan2017axiomatic}—attribute individual predictions to specific input tokens. While useful for debugging, these explanations are often shallow and non-generalizable, particularly in source code, where tokens are discrete, densely structured, and context-sensitive. For example, attribution methods may highlight tokens like \texttt{<} without indicating whether the model treats them as comparison operators, type delimiters, or frequent lexical patterns. 

Global methods, such as probing classifiers~\cite{karmakar2021pre, troshin2022probing, wan2022they, lopez2022ast, 10.1145/3691621.3694931}, assess whether latent representations encode properties such as syntax or data types. While effective in some domains, probing relies on predefined labels and assumptions about what should be encoded. In code, the deterministic and repetitive nature of syntax can inflate probing accuracy without offering true insights into abstraction~\cite{belinkov2022probing, 10.1145/3691621.3694931}. As such, both local and global methods leave critical gaps in understanding how models conceptualize programs. 

This gap motivates a shift toward a complementary \textit{model-centric} interpretability—approaches that discover meaningful abstractions from model representations without relying on handcrafted features or provide shallow explanations. Further, recent work in NLP and computer vision reflects a growing shift toward concept-based interpretability, including Concept Bottleneck Models~\cite{sun2024cbllm}, concept-aware language models~\cite{shani2023towards, liao2023concept}, and concept-based explanations~\cite{yeh2022humancenteredconceptexplanationsneural}. These efforts underscore the value of aligning internal representations with semantically meaningful abstractions and have led to structured resources such as D-Concept~\cite{liao2023concept}, a dataset for the hypernym discovery task to evaluate LLMs’ ability to distinguish between abstract and concrete concepts. However, such developments remain unexplored in the code domain.

We introduce Code Concept Analysis (CoCoA), a global post-hoc interpretability framework for code models which clusters contextualized token activations to uncover emergent structure in the model’s representation space, revealing its organization of lexical, syntactic, and semantic concepts without relying on hand-crafted features or task-specific supervision. We annotate these concepts using alignment with pre-defined concepts and LLM-as-an-annotator and construct CodeConceptNet (CoCoNet) dataset.  We demonstrate that CoCoA can enhance local explanations by mapping salient tokens to concept clusters, yielding more faithful and human-interpretable explanations. For example, CoCoA disambiguates whether a highlighted \texttt{<} token functions as a comparator, a generic type delimiter, or a PHP opening tag (\texttt{<?}).

\vspace{1mm}
\noindent
\textbf{Our contributions are as follows:}
\begin{enumerate}
    \item We introduce \textbf{Code Concept Analysis (CoCoA)}, a global post-hoc interpretability framework that clusters contextualized token representations to uncover emergent latent concepts in code language models.

    \item We annotate concept clusters along lexical, syntactic, and semantic dimensions using a combination of predefined tags and LLM-generated descriptions, without relying on task-specific supervision.

    \item We analyze how latent concepts evolve across layers, shift under fine-tuning, and respond to semantic-preserving perturbations.

    \item We show how latent concepts can enhance local attributions by mapping salient tokens to concept clusters.
\end{enumerate}

\noindent
The remainder of the paper is organized as follows: Section~\ref{Methodology} describes our methodology; Section~\ref{Evaluation} presents our evaluation; Section~\ref{Related Work} reviews related work; Section~\ref{Limitations} outlines limitations, and Section~\ref{Conclusion} concludes the paper.

\section{Methodology}
\label{Methodology}

Our \textbf{Code Concept Analysis (CoCoA)} pipeline comprises the following components: (1) Concept Discovery, (2) Concept Alignment and Annotation, and  (3) an application of Latent Concept Attribution. 

\begin{figure*}[t]
    \centering
    \includegraphics[width=\textwidth]{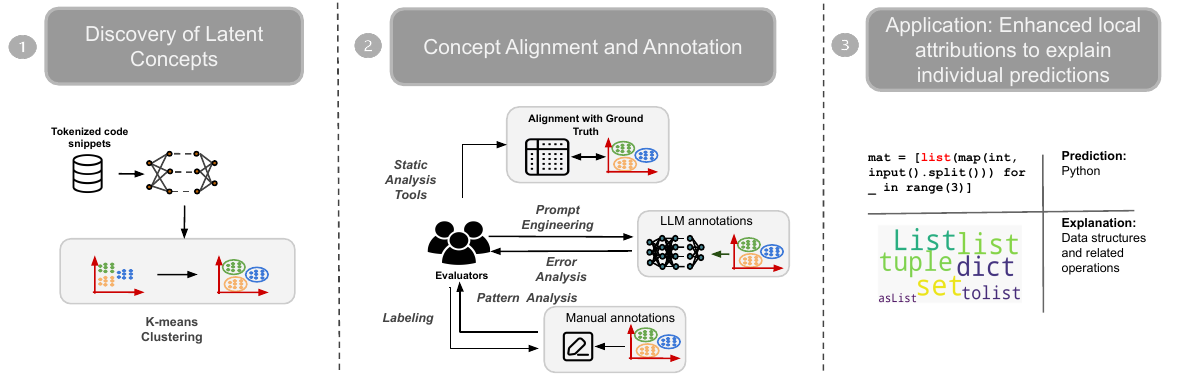}
    \caption{Workflow of CoCoA methodology.}
    \label{fig:evaluation framework}
\end{figure*}

\subsection{Concept Discovery}
\label{Methodology: Concept Discovery}
We perform Latent Concept Analysis ~\cite{alam2022conceptxframeworklatentconcept,hawasly2024scaling} on a model's contextualized representations. Let $\mathbb{M}$ denote a pre-trained code language model that maps an input $x$ (e.g., a code snippet) to a contextualized token representation $\mathbb{M}(x) \in \mathbb{R}^d$, where $d$ is the hidden dimension. These representations capture rich, layer-wise information reflecting both surface form and higher-level semantics. To discover latent concepts, we extract activations from each model layer and apply K-Means clustering to group token embeddings based on representational similarity. K-Means partitions the data to minimize intra-cluster variance, iteratively assigning points to centroids until convergence. Each resulting cluster reflects a recurring pattern in the model’s learned abstraction space.

\subsection{Concept Alignment and Annotation}

We annotate clusters using two strategies: (1) alignment with predefined syntactic labels, and (2) discovery of novel semantic categories using human annotation and LLMs.

\paragraph{Alignment with Ground-Truth.} 
We first analyze lexical patterns such as affixes, casing, and n-grams. This is followed by syntactic evaluation, where we use Tree-sitter-generated AST labels to measure alignment.

Let \( \mathcal{C}_E = \{C_e\} \) denote the set of discovered clusters and \( \mathcal{C}_H = \{C_h\} \) denote the set of ground-truth categories, where each \( C_e, C_h \subseteq \mathcal{T} \), the full token set.

\textit{Alignment} \( \alpha_\theta(C_e) \): A discovered cluster \( C_e \) is said to align with a ground-truth category \( C_h \) if it overlaps significantly with at least one \( C_h \in \mathcal{C}_H \), exceeding a token overlap threshold \( \theta \):
\[
\alpha_\theta(C_e) = 
\begin{cases}
1, & \text{if } \exists\ C_h \in \mathcal{C}_H : \frac{|C_e \cap C_h|}{|C_e|} \geq \theta \\
0, & \text{otherwise}
\end{cases}.
\]

\textit{Coverage} \( \kappa_\theta(C_h) \): A ground-truth category \( C_h \) is considered covered if it is captured by at least one discovered cluster \( C_e \in \mathcal{C}_E \):
\[
\kappa_\theta(C_h) = 
\begin{cases}
1, & \text{if } \exists\ C_e \in \mathcal{C}_E : \frac{|C_e \cap C_h|}{|C_e|} \geq \theta \\
0, & \text{otherwise}
\end{cases}.
\]

\paragraph{Manual Annotation and LLM-as-Annotator.}
In the absence of labeled semantic ground-truth datasets, we initially performed manual annotation with the help of computer science senior undergraduates with Java experience. Annotators labeled clusters using token lists and example code contexts, resulting in 500 annotated clusters. 

However, manual annotation is labor-intensive and difficult to scale. To address this, we used the \texttt{GPT-4o} model in a zero-shot setting to generate cluster labels and descriptions. Early prompts excluded context due to rate limits, resulting in inaccuracies (e.g., labeling \texttt{<} as an “Opening Bracket” rather than a type constraint). To improve reliability, we developed a few-shot prompt based on annotated examples and added clarifications to disambiguate common edge cases (e.g., function calls vs. grouping parentheses). Figure \ref{fig:evaluation framework} shows one such few shot example. Model outputs followed a standardized JSON format with fields for label, semantic tags, and description, enabling consistency in downstream evaluation. Prompt design was refined through manual error analysis.

From 1,724 raw labels, we consolidated a curated set of 43 canonical semantic tags through manual filtering and frequency analysis. Subsequent evaluations showed most clusters aligned with this limited tag set with only a few corresponding to Unclear Behavioral Role label. We conducted a user study across 500 clusters comparing LLM-generated labels to human annotations and found that LLMs often produced more consistent and semantically expressive tags (section \ref{Module Impact Study}).

\subsection{Concept-based Local Explanations}
\label{sec:lacoat_methodology}

In this section, we present an application of Latent Concept Analysis by integrating it with local attribution methods such as Integrated Gradients. Traditional attribution methods highlight individual tokens that influence model predictions, but prior work suggests that models often over-rely on syntactic markers, resulting in shallow explanations with limited semantic grounding. These methods offer little insight into whether the attributions reflect meaningful abstractions or spurious correlations. To address this, we extend attribution from individual tokens to latent concepts—clusters of contextualized representations—thereby elevating explanations to a higher level of abstraction. Adapting techniques from recent work in NLP~\cite{yu2024latent}, we apply this framework to code-related tasks such as error detection and language classification. We investigate whether mapping salient tokens to latent concepts yields more robust and interpretable explanations. These concepts reveal the model’s internal organization and serve as high-level, semantically coherent explanations of its predictions.

Formally, the process has two steps: (1) Given an input instance $s$ and a prediction $p$ from model $\mathcal{M}$, we first identify and interpret the salient internal representations that contribute to the prediction. We first apply Integrated Gradients~\cite{sundararajan2017axiomatic} to compute attribution scores over input tokens. We then select the top tokens whose cumulative attribution accounts for 50\% of the total attribution mass, following a top-$P$ sampling strategy. 
(2) Each salient token representation is then mapped to a latent concept $C_i$ previously discovered by during the Concept Discovery (\cref{Methodology: Concept Discovery}). To perform this mapping, we train a lightweight logistic regression classifier that maps each token representation $\vec{z}_{w_i}$ to one of $K$ latent concepts. 
The classifier is trained using the representations obtained from the training data $\mathcal{D}$, where each input feature is the contextualized representation of a token $\vec{z}_{w_j}$, and the corresponding label is the index $i$ of the concept cluster $C_i$ to which the token belongs. That is, for every token $w_j \in C_i$, the training pair is $(x = \vec{z}_{w_j},\ y = i)$. 

We further conduct a user study to validate the usefulness of our method in enhancing local attribution-based explanations. The study can be found in section \ref{Module Impact Study}.






\section{Evaluation and Discussion}
\label{Evaluation}

\subsection{Experimental Settings}
\label{Experimental Settings}

\paragraph{Datasets and Models} We randomly select 30k code snippets from the CodeNet dataset~\cite{puri2021codenet}. We first tokenize the sentences using the Tree-sitter library
and pass them through the
standard pipeline of CodeBERT as implemented in HuggingFace. We extract layer-wise representations
 of every token using the NeuroX library\footnote{\url{https://github.com/fdalvi/NeuroX}}. We perform Latent Concept Analysis on three models CodeBERT\cite{feng2020codebert}, UniXCoder \cite{guo2022unixcoder} and DeepSeekCoder-V2-Lite \cite{zhu2024deepseek}. We discuss dataset and model selection in  appendix \ref{Appendix: Model and dataset choices} and generalizability across models in section \ref{Generalization across models}. We use the huggingface transformers \footnote{\url{https://huggingface.co/}} library with the default settings and hyperparameters.

For Latent Concept Attribution, we finetune the CodeBERT model on three tasks created from the CodeNet dataset: AST token tagging, compilation error detection and programming language classification tasks. AST Token Tagging is a sequence
labeling task, while the other tasks are sequence
classification tasks. Details of all the datasets are provided in the appendix \ref{Dataset details}. 

\paragraph{Clustering} To optimize clustering quality, we limit the vocabulary to approximately 310k tokens by discarding those with a frequency above 15K. Frequent tokens (e.g., \texttt{int}, \texttt{\{}, \texttt{;}) tend to dominate clusters, leading to large, uninformative groupings. We empirically set the number of clusters to $K = 350$. We further discard extremely large clusters (e.g., those containing over 15K tokens), as they obscure structural nuances critical to understanding model behavior. \footnote{The elbow method was considered for automatic selection of $K$, we did not observe a well-defined inflection point across layers.  We adapt the procedure laid out by \citet{dalvi2022discovering}.}

\paragraph{LLM settings for annotation} We used Gemini-2.0-Flash cite for annotating the clusters.
We chose a temperature of 0.2, a top-p setting of 0.4 and top-k of 8  to
optimize for reliable, high-quality annotations using a dictionary of semantic labels.

\subsection{Latent Concept Analysis and Evolution}

We analyze latent concept clusters from the final layers of the models across the pretraining and three fine-tuning objectives: AST node classification, compile error detection, and language classification. We find that discovered clusters often align with human-interpretable categories across three abstraction levels:
\textbf{Lexical} clusters group tokens with surface-level similarities (e.g., \texttt{tab1}, \texttt{sum1}, \texttt{ans1} for variable names ending in “1” or numerics like \texttt{100}, \texttt{26}); 
\textbf{Syntactic} clusters include structural roles such as \texttt{public}, \texttt{private}, \texttt{protected} (access modifiers), or comparison operators like \texttt{<}, \texttt{==}, \texttt{!=};
\textbf{Semantic} clusters reflect functional behavior or intent—e.g., \texttt{abs}, \texttt{max}, \texttt{floor}, and \texttt{pow} forming a “mathematical operations” cluster, or tokens like \texttt{logger.error} indicating error reporting logic. We provide the dataset with our clusters and annotations for further interpretability studies in supplementary material. 

\begin{figure*}[htbp]
    \centering
C    \includegraphics[width=\textwidth]{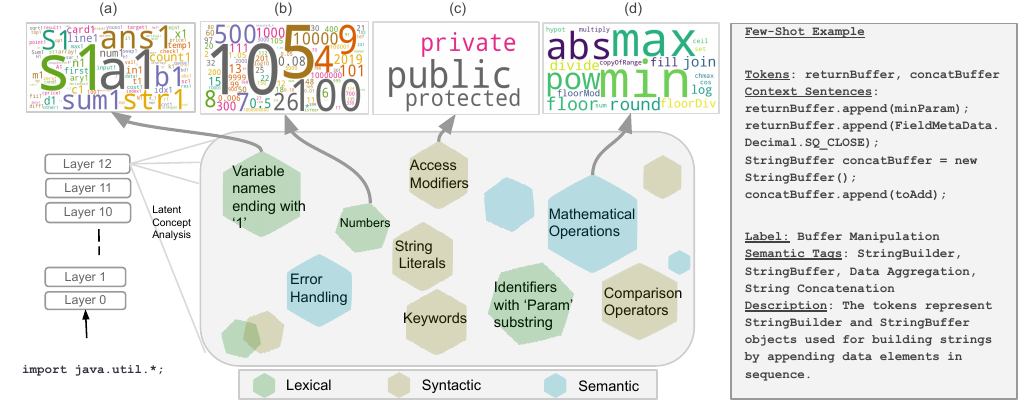}
    \caption{Lexical, syntactic and semantic clusters found in model latent representations. See Appendix~\ref{Appendix: Full_prompt} for the full prompt.}
    \label{fig:evaluation framework}
\end{figure*}

Table~\ref{tab:lexical_patterns} reports lexical patterns such as affixes, casing, and n-grams, while Table~\ref{tab:alignment_metrics_codenet_treesitter} quantifies alignment with AST-derived syntactic labels using alignment and coverage metrics. A cluster is considered aligned if at least 90\% of its tokens share a predefined syntactic label, including syntactic categories. 

\paragraph{Emergent concepts and a model-induced taxonomy:}
Building on these clusters, we introduce a novel dataset of semantic labels derived not from external ontologies but from the model’s own internal organization. Labels are assigned post hoc using LLM-guided annotation of latent clusters. This approach not only reveals new categories of functional behavior, but also enables scalable evaluation of internal conceptual structure in code models. Unlike NLP, the code domain lacks widely accepted semantic evaluation sets. \textbf{Code ConceptNet (CoCoNet)} represents a first step toward filling this gap with a dataset grounded in emergent model behavior. It includes clusters such as those involving mathematical operations, Python data structures and associated methods, or function definition keywords across languages (e.g., \texttt{function} in Javascript and \texttt{def} in Python).


\paragraph{Evolution of concepts:}
While prior work broadly characterizes a layer-wise transition from lexical to semantic features, our analysis offers a more granular perspective on how these abstractions emerge and interact in code language models. Lexical alignment is most pronounced in the lower layers, particularly for features such as prefixes, suffixes, and substring patterns. These are often artifacts of subword tokenization schemes but are nonetheless reflected in the model’s learned representations. Code Concept Analysis reveals that many clusters in these early layers are formed around shared lexical properties. Notably, lexical patterns persist into deeper layers, but are increasingly embedded within structurally coherent clusters—such as identifier suffixes. This suggests that the model integrates surface-level lexical cues into more abstract representations over depth. We investigate this further in Section~\ref{sec:evaluation_robustness} through lexical perturbation experiments, which test whether the model overrelies on superficial lexical features. Our findings indicate that, although lexical traits (e.g., naming conventions for variables or classes) continue to influence representation space, they are often nested within higher-order syntactic structures. This points to the model’s capacity to encode generalizable representations that transcend purely lexical regularities, even under the constraints of subword segmentation.

\begin{table*}[ht]
  \caption{Lexical alignment of token clusters from CodeBERT layer 12 (\textbf{350 clusters}) across pretraining and fine-tuning stages. Evaluated at 80\% similarity threshold.}
  \label{tab:lexical_patterns}
  \centering
  \scriptsize
  \begin{tabular}{lcccc}
    \toprule
    \textbf{Lexical Pattern (\%)} 
    & \textbf{Pre-trained} 
    & \textbf{AST Node Classification} 
    & \textbf{Compile Error Detection} 
    & \textbf{Language Classification} \\
    \midrule
    Substring match (>3) & 21.7 & 26.0 & 14.9 & 3.4 \\
    Prefix                & 32.9 & 35.1 & 37.1 & 12.3 \\
    Suffix                & 30.6 & 34.3 & 34.0 & 9.4 \\
    \midrule
    Camel Casing          & 4.0  & 12.0 & 3.7  & 0.3 \\
    Pascal Casing         & 8.3  & 5.1  & 10.6 & 1.4 \\
    \bottomrule
  \end{tabular}
\end{table*}

\begin{table*}[ht]
  \caption{Syntactic alignment metrics for CodeBERT evaluated at varying thresholds. For language classification, Tree-sitter tags are drawn from all six task languages.}
  \label{tab:alignment_metrics_codenet_treesitter}
  \centering
  \scriptsize
  \setlength{\tabcolsep}{2pt}
  \begin{tabular}{p{3.2cm}ccc|ccc|ccc|ccc}
    \toprule
    \textbf{Metric} 
    & \multicolumn{3}{c|}{\textbf{Pre-trained}} 
    & \multicolumn{3}{c|}{\textbf{AST Node Classification}} 
    & \multicolumn{3}{c|}{\textbf{Compile Error Detection}} 
    & \multicolumn{3}{c}{\textbf{Language Classification}} \\
    \cmidrule(r){2-4} \cmidrule(r){5-7} \cmidrule(r){8-10} \cmidrule(r){11-13}
    & 85\% & 90\% & 95\%
    & 85\% & 90\% & 95\%
    & 85\% & 90\% & 95\%
    & 85\% & 90\% & 95\% \\
    \midrule
    Clusters Labeled (/350)       & 332 & 328 & 314 & 342 & 341 & 338 & 312 & 306 & 292 & 308 & 295 & 273 \\
    Tag Coverage (\%)             & 50.0 & 50.0 & 50.0 & 76.4 & 76.4 & 75.0 & 37.5 & 34.7 & 31.9 & 12.7 & 12.7 & 12.7 \\
    Overall Alignment Score       & 0.724 & 0.719 & 0.699 & 0.871 & 0.869 & 0.858 & 0.633 & 0.611 & 0.577 & 0.504 & 0.485 & 0.454 \\
    Unique Tags Identified        & 36 & 36 & 36 & 55 & 55 & 54 & 27 & 25 & 23 & 21 & 21 & 21 \\
    Unaligned Clusters            & 18 & 22 & 36 & 8  & 9  & 12  & 38 & 44 & 58 & 42 & 55 & 77 \\
    \bottomrule
  \end{tabular}
\end{table*}

\begin{figure}[h]
  \centering
  \includegraphics[width=0.95\textwidth]{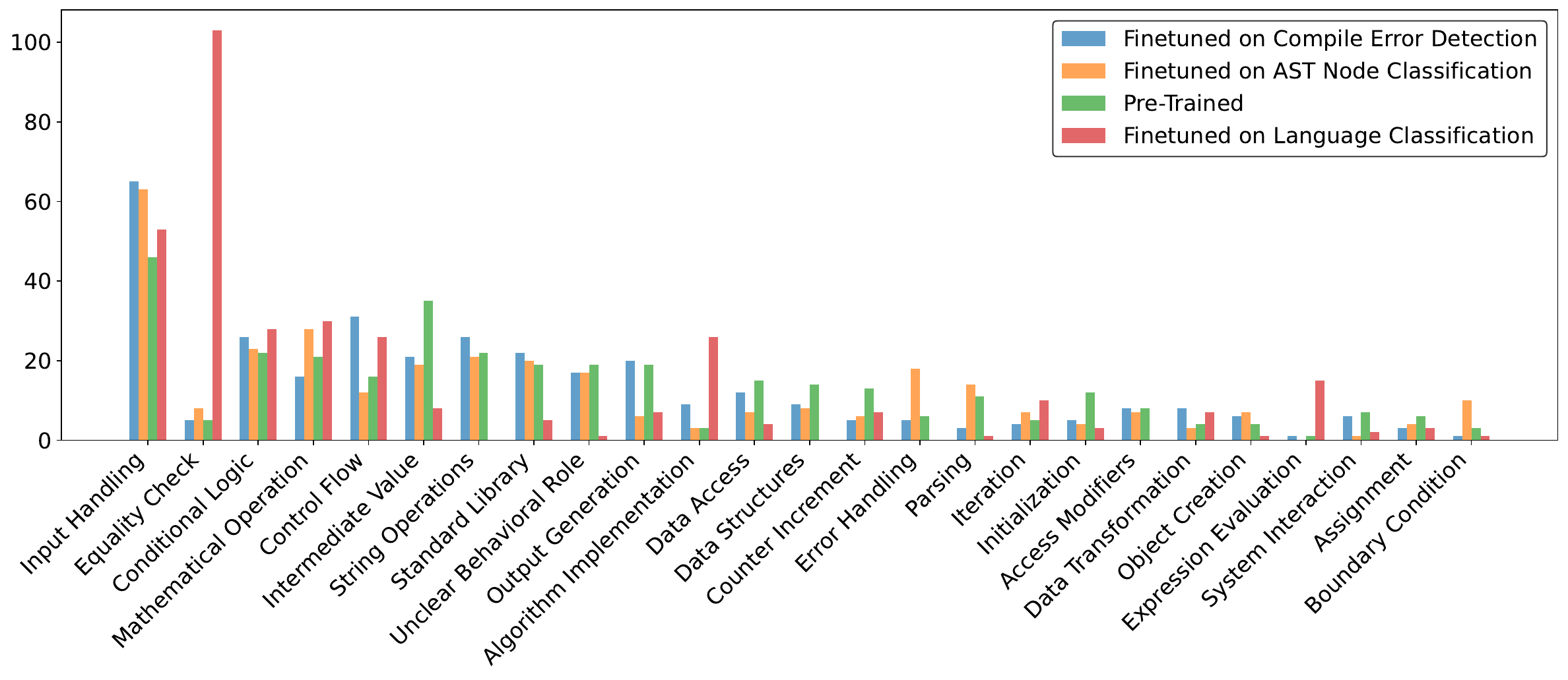}
  \caption{Top 25 Semantic Cluster Labels for CodeBERT. Complete list can be found in the supplementary material }
  \label{fig:semantic tags}
\end{figure}


Syntactic alignment is highest in the early and middle layers and either plateaus or declines slightly in deeper layers. A full inventory of the syntactic tags used in our analysis is included in the supplementary material.Figure~\ref{fig:semantic tags} displays the most frequent semantic tags across clusters. Notably, we identify 21 clusters that could not be readily aligned with predefined roles and were labeled as having an \texttt{Unclear Behavioral Role}.

Our analysis reveals the presence of \textit{compositional latent clusters}—groups of tokens that blend fine-grained concepts such as control flow, error handling, and identifier usage, without aligning cleanly to any single syntactic or semantic tag. These clusters appear to capture interactions between multiple latent features, reflecting higher-order abstractions learned by the model which might be useful for generalization. Probing these compositions at higher levels of granularity is an important future work direction.

\paragraph{Effect of Fine-Tuning.}
Fine-tuning on structure-aware objectives, such as AST node classification, increases the proportion of clusters reflecting lexical regularities and structural patterns. This suggests that such tasks promote the learning of syntax-sensitive features. We also observe a clear reduction in unaligned clusters, indicating improved syntactic alignment and a more coherent internal organization. These trends align with prior research findings that structural supervision enhances syntactic understanding. Conversely, fine-tuning on semantically broader tasks—such as compile error detection and language classification—leads to a modest decline in lexical and syntactic alignment, reflecting a shift toward higher-level or compositional abstractions. This is most pronounced for language classification, where unaligned groups drop to just 2 to 3 (figure \ref{fig:semantic tags}). Manual inspection shows that these clusters often encode abstract patterns that generalize across languages, suggesting the emergence of cross-lingual conceptual groupings, potentially at the expense of syntactic precision.

\paragraph{Generalization across models.}
\label{Generalization across models}
Despite differences in architectures, training objectives, and capacities, we observe consistent patterns in the evolution of concept alignment and abstraction across all three models. UniXCoder exhibits a more pronounced drop than the gradual decline observed in CodeBERT and DeepSeekCoder. But the overall decreasing trend suggests that deeper layers increasingly prioritize task-specific, semantic, or compositional abstractions over structural or syntactic signals—a pattern consistent with prior observations in transformer-based architectures, where late-stage representations often emphasize functional over formal structure.

Across models, we find that discovered clusters consistently map onto a small, shared set of semantic concepts. The top 25 semantic labels account for the vast majority of cluster alignments in all three models, indicating strong conceptual convergence despite architectural and training differences (Figure~\ref{fig:semantic tags}). Detailed results for UniXCoder and DeepSeekCoder, including layer-wise syntactic alignment metrics and semantic label distributions, are provided in the supplementary material.

\begin{figure}[h]
    \centering

    \subcaptionbox[CodeBERT pre-trained]{\scriptsize CodeBERT  \label{fig:codebert_codenet4k_ast_lex}}{
        \includegraphics[width=0.30\textwidth]{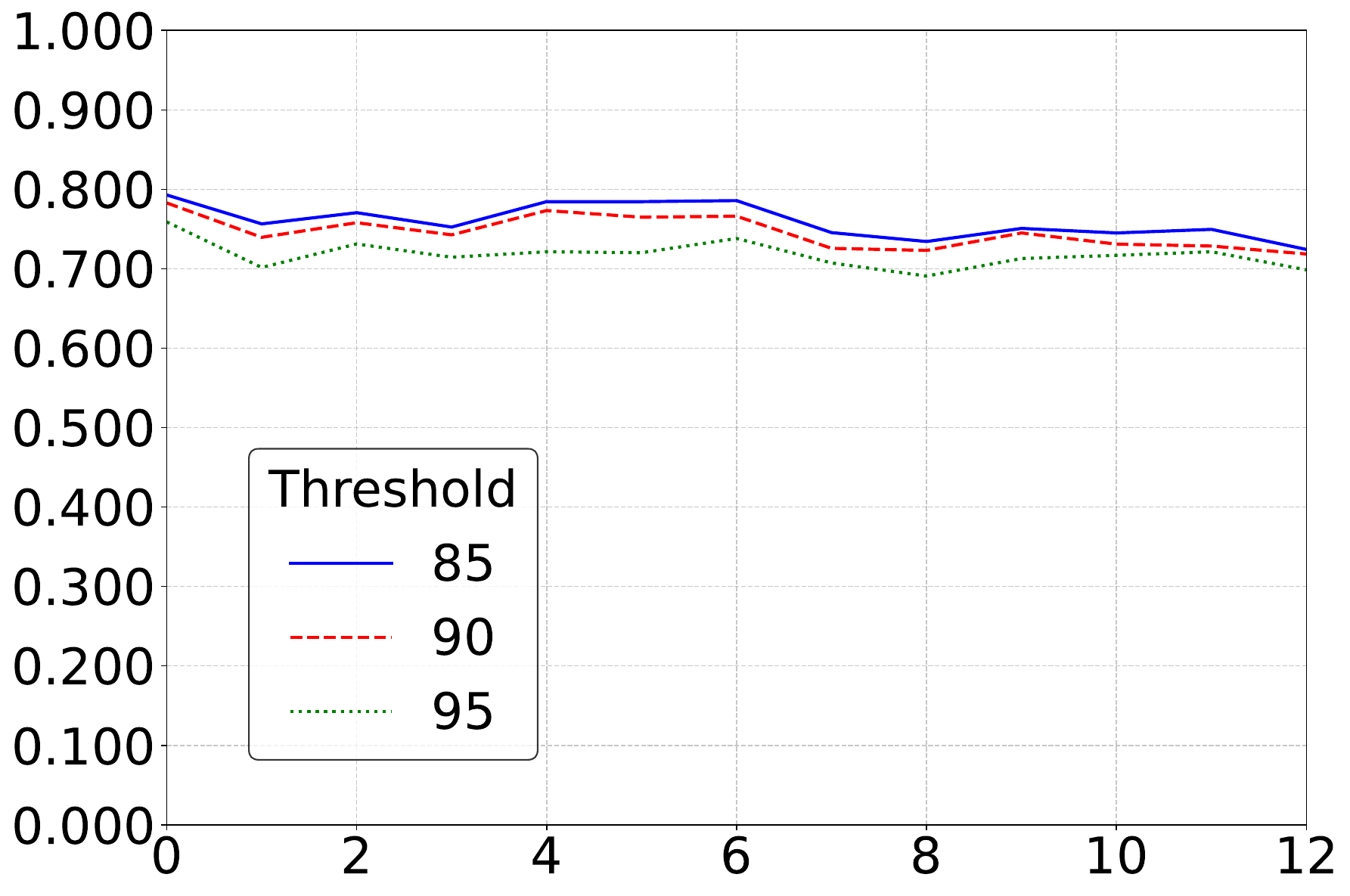}
    }
    \hspace{-0.3cm}
    \subcaptionbox[UnixCoder pre-trained]{\scriptsize UnixCoder  \label{fig:unixcoder_codenet4k_ast_lex}}{
        \includegraphics[width=0.30\textwidth]{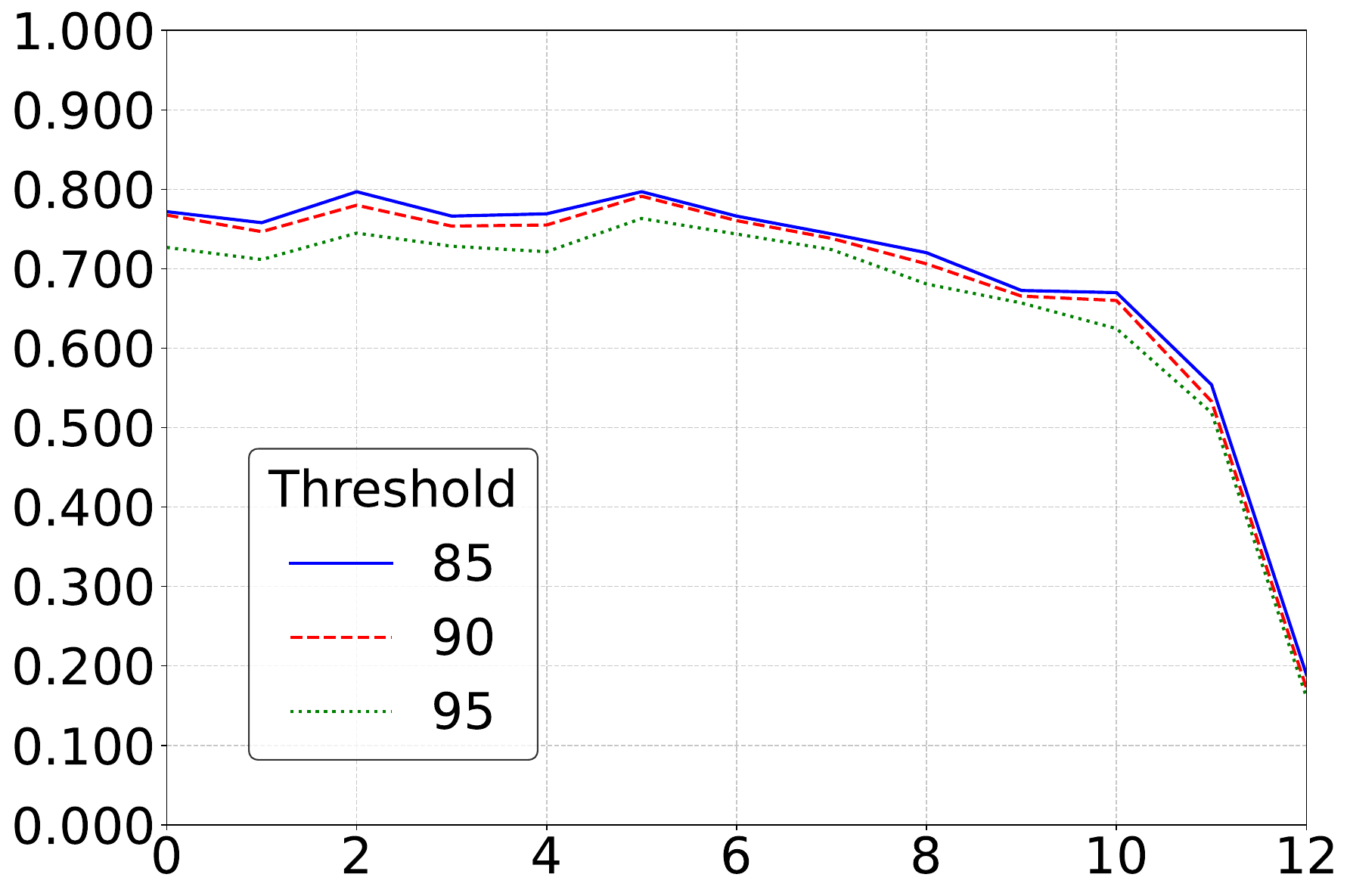}
    }
    \hspace{-0.3cm}
    \subcaptionbox[DeepSeekCoder pretrained]{\scriptsize DeepSeekCoder\label{fig:codebert_ast_finetuned_lex}}{
        \includegraphics[width=0.30\textwidth]{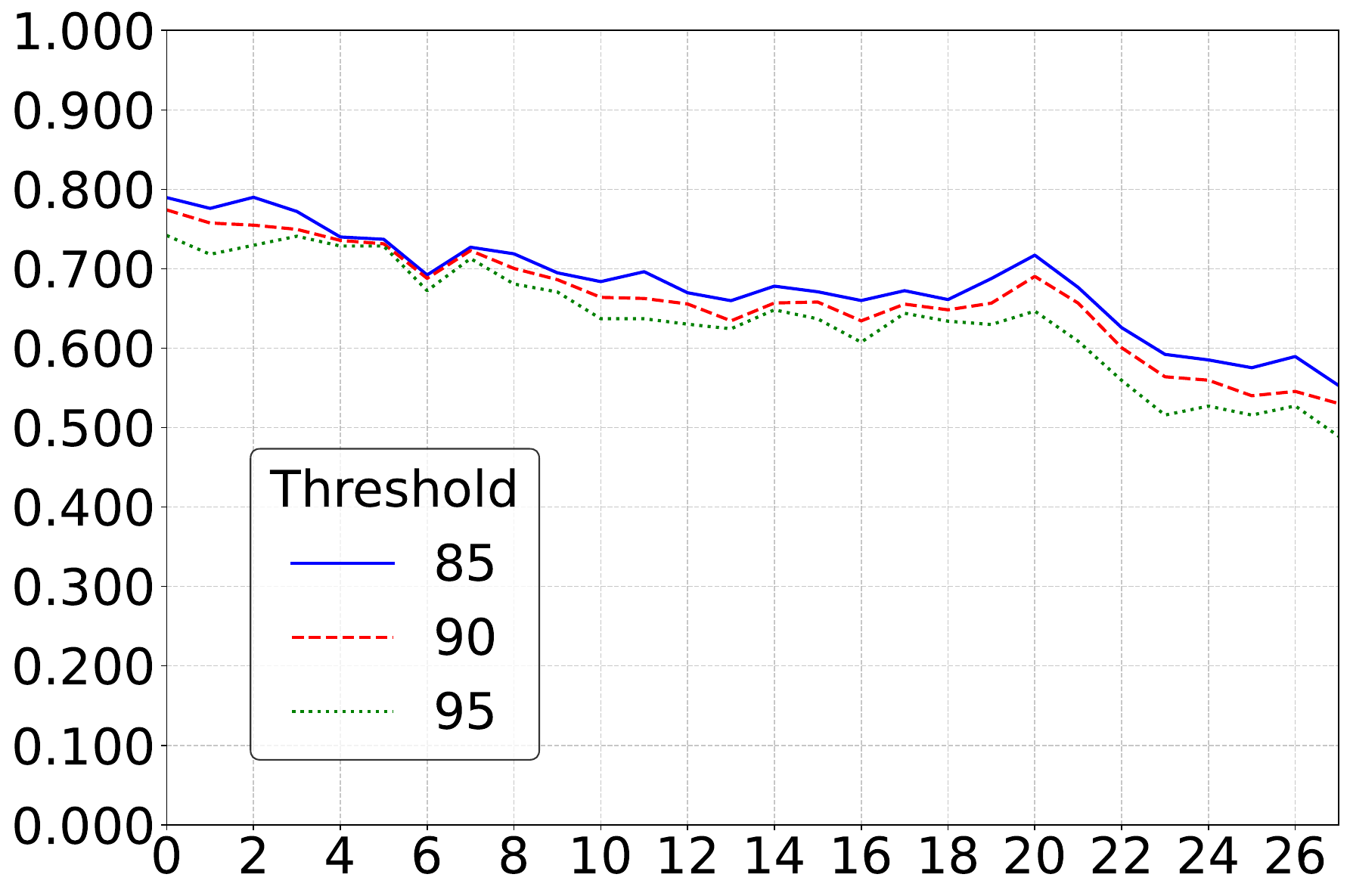}
    }

    \caption{
    Syntactic alignment across three models
    }
    \label{fig:alignment_scores_by_model_and_task}
\end{figure}

\subsection{Robustness Under Semantic-Preserving Perturbations}
\label{sec:evaluation_robustness}

We assess the stability of latent concept clusters under semantic-preserving code transformations using the Cluster Sensitivity Index (CSI), defined as \(\text{CSI} = 1 - \text{Average Jaccard Similarity}\), where lower values indicate greater robustness (Appendix~\ref{Jaccard}).

Lexical perturbations—such as identifier renaming((CSI $\approx$ 0.33), case changes, and canonical substitutions—preserve surface form while maintaining semantics. Structural perturbations, including scope reassignment and statement reordering, result in greater disruption (CSI up to 0.47), with statement order randomization being most impactful. These results show that LCA-derived clusters are semantically grounded and robust to superficial changes—unlike token-level saliency methods, which often fluctuate under such edits. This robustness supports their use in reliable, concept-based model explanations. This motivates our application in the following section.

\subsection{Application: Latent Concept Attribution}
\label{Application: LACOAT}
Local feature attribution methods, such as Integrated Gradients, highlight individual tokens that contribute most to a model’s prediction—for example, predicting the programming language of a given code snippet. However, these salient tokens often do not align with human intuition. Prior studies have shown that code LLMs frequently rely on superficial cues, such as punctuation or formatting, rather than deeper semantic content~\citep{rabin2021understanding, zhang2022diet}.

In this work, we investigate whether the model's reliance on such salient tokens reflects true conceptual understanding or merely spurious correlations. To this end, we train a logistic regression classifier to map attribution-derived salient tokens to latent concept clusters discovered via unsupervised clustering of contextualized token representations. We then use an LLM to generate human-understandable explanations for the important tokens and their associated clusters. 

\begin{figure*}[ht]
    \centering
    \includegraphics[width=\textwidth]{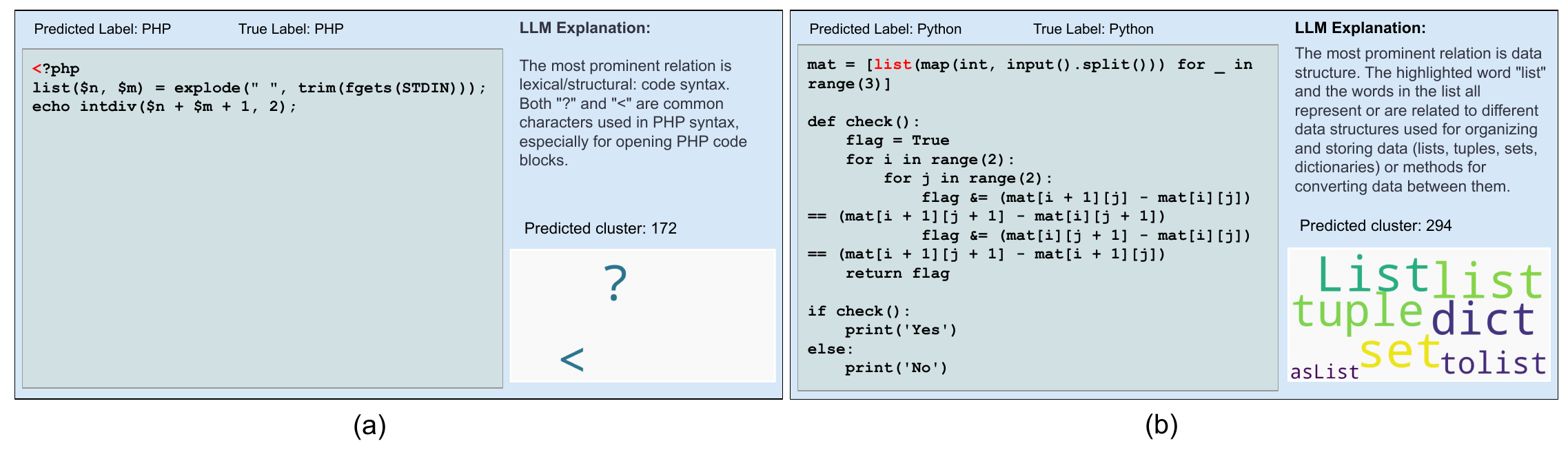}
    \caption{Examples of Latent Concept Attribution applied to a programming language classification task. Each example shows a code snippet with its predicted and true label, the top salient tokens selected via attribution, the predicted concept cluster, an LLM generated explanation.}

    \label{fig:lacoat}
\end{figure*}

Our results suggest that, in many cases, these salient tokens correspond to semantically coherent latent concepts rather than isolated artifacts. For instance, if a token such as \texttt{<} is deemed important for a prediction, we examine the latent concept cluster it belongs to. If this cluster predominantly contains other logical operators (e.g., \texttt{>}, \texttt{==}, \texttt{!=}), this suggests that the model has abstracted a higher-level concept corresponding to logical comparison or in case of \ref{fig:lacoat} \texttt{<} is for PHP opening blocks (<?) and not a logical comparison as we can tell from the cluster. This indicates that the important tokens are selected based on meaningful, generalizable features—structured interactions embedded within the representation space—that go beyond surface-level patterns. These latent interactions, while not always aligned with explicit syntax, reflect non-trivial model behavior and warrant further study as a basis for robust and interpretable predictions.

\subsection{Impact Study}
\label{Module Impact Study}

\paragraph{Impact of prompt engineering on quality of labels}
We conducted a manual evaluation to assess the quality of LLM-generated labels produced by GPT-4o, focusing on the effect of prompt engineering. With tailored prompts, 95.8\% of the generated labels were rated as acceptable by human annotators. In direct comparison with manual annotations, 38.4\% of the syntactic labels were judged to be superior, while 83.2\% of the semantic labels were deemed more accurate and informative than their human-generated counterparts. These results demonstrate that prompt engineering not only improves label acceptability but can also yield annotations that surpass human baselines in semantic clarity. Annotation details are provided in appendix~\ref{Annotation Guidelines}.

\paragraph{Impact of latent concept attributions on local explanations}
To further understand the utility of latent concept attribution, we conducted a user study analyzing how well individual predictions are explained with and without concept-based enhancements. We found that only 24.3\% of the clusters could be satisfactorily explained using token-level attributions alone, such as those derived from integrated gradients. In contrast, an additional 37.0\% of the clusters benefited from incorporating latent concept information—these were cases where the salient token alone was insufficient, but the broader concept cluster provided a coherent and useful explanation. This finding highlights the complementary nature of latent concept attribution in improving the semantic plausibility and interpretability of local explanations.

To assess the consistency of judgments made by human evaluators, we computed the \textbf{Inter-annotator agreement} using Fleiss' $\kappa$ statistics across two key questions in the annotation study. We observe moderate inter-annotator agreement (table~\ref{tab:fleiss_kappa}), suggesting reasonably consistent human judgments despite the subjectivity of the task. Appendix \ref{Annotation Guidelines} provides annotation details. 

\begin{table}[h]
  \caption{Fleiss' $\kappa$ evaluation of annotator agreement on binary labeling questions.}
  \label{tab:fleiss_kappa}
  \centering
  \begin{tabular}{p{7.2cm}cc}
    \toprule
    \textbf{Question} & \textbf{Yes (\%)} & \textbf{Fleiss' $\kappa$} \\
    \midrule
    Q1: Does the salient token explain prediction? &  24.3\% & 0.577 \\
    Q2: Does cluster context help explain prediction?        & 48.6\% & 0.525 \\
    \bottomrule
  \end{tabular}
\end{table}

\section{Related Work}
\label{Related Work}

Recent work has approached the explainability of code language models from diverse angles. A number of methods focus on identifying influential input tokens for individual predictions. For example, WheaCha~\cite{wang2023explanation} partitions tokens into “wheat” (predictive features) and “chaff” (non-essential context), revealing that model decisions often hinge on shallow lexical or syntactic cues. Attribution techniques such as SHAP and Integrated Gradients have also been applied to code~\cite{cito2022counterfactual}, though they frequently fail to yield semantically coherent explanations. Syntax-grounded approaches like AST-Probe~\cite{hernandez2022astprobe} and ASTrust~\cite{palacio2024towards} map internal representations to Abstract Syntax Tree (AST) structures, aligning model confidence with human-defined syntactic categories. While ASTrust supports both local and global explanations, it remains dependent on predefined grammar constructs. Similarly, DeciX~\cite{10.1145/3660814} leverages causal inference to quantify the influence of token dependencies, further underscoring the model’s reliance on syntactic reasoning. CodeQ~\cite{palacio2025explaining} introduces the notion of code rationales—minimal token subsets responsible for predictions—which are aggregated to form dataset-level explanations. However, CodeQ operates solely at the input level, without addressing the latent structures learned during model training.Other methods have explored counterfactual explanations~\cite{cito2022counterfactual}, perturbing source code to identify minimal edits that change predictions. While valuable, these techniques typically assume fixed interpretability primitives and remain local in scope. Attention-based methods are not considered reliable explanations \cite{jain2019attention,wiegreffe2019attention}.  In contrast, our approach uncovers and annotates latent concepts—emergent clusters of token representations within the model’s hidden space. 



\section{Limitations}
\label{Limitations}

(1) Concept discovery over contextualized representations involves a trade-off between granularity and scalability. While agglomerative clustering captures fine-grained hierarchical structure, its high computational cost—particularly due to full dendrogram construction—limits scalability. As a result, we applied it only to smaller subsets of the data, reducing coverage and diversity. We instead adopt K-Means, which scales better but assumes spherical clusters and may miss irregular boundaries. Prior work~\cite{hawasly2024scaling} suggests K-Means can yield comparably meaningful concepts in NLP models, but exploring more scalable alternatives for code-specific representations remains a promising direction.
(2) While our latent concept framework improves the interpretability of local attributions, it inherits the limitations of attribution methods themselves—particularly their tendency to focus on syntactic tokens. Moreover, we evaluated our approach on two tasks: programming language classification and compile error detection. The latter posed challenges due to attribution instability in complex semantic contexts. However, our user study shows that concept-enhanced explanations can still offer value for simpler tasks like language classification.

\section{Conclusion and Future Work}
\label{Conclusion}

We introduced a framework for interpreting code language models through Latent Concept Analysis (LCA), which clusters hidden representations to reveal lexical, syntactic, and semantic abstractions. Our analysis shows that lexical patterns—shaped by subword tokenization—persist in deeper layers, but are embedded within structurally coherent contexts. Fine-tuning reshapes these abstractions in task-specific ways: structure-aware objectives sharpen syntactic boundaries, while semantic tasks promote cross-lingual generalizations. These trends are consistent across models and datasets, with many high-level semantic concepts recurring across model variants.

To support scalable concept interpretation, we proposed an LLM-guided annotation strategy and introduced CodeConceptNet (CoCoNet), a dataset that grounds cluster semantics in model-internal behavior. This resource complements prior work in NLP (e.g., D-Concept~\cite{liao2023concept}) by establishing structured, interpretable labels tailored to code. We further demonstrated how latent concepts can enhance local explanation methods. By aligning saliency with semantically coherent clusters, our approach improves interpretability—as validated in a user study.

In future work, we plan to move beyond token-level attribution and focus on concept-level model interpretation, including concept-aware training objectives that may improve both interpretability and robustness.




\bibliographystyle{ACM-Reference-Format}
\bibliography{references}

\newpage
\appendix
\section*{Appendix}
\section{Model and Dataset Selection}
\label{Appendix: Model and dataset choices}

\subsection{Models}

To evaluate the generality of our framework, we analyze three widely used pretrained models with diverse architectures, training objectives, and scales: CodeBERT~\cite{feng2020codebert}, UniXCoder~\cite{guo2022unixcoder}, and DeepSeekCoder V2 Lite.\footnote{We use the \texttt{DeepSeek-Coder-V2-Lite-Instruct} variant.}

CodeBERT is an encoder-only transformer pretrained on paired code and natural language using masked language modeling and replaced token detection, serving as a standard baseline for code understanding.

UniXCoder extends this architecture with additional pretraining tasks, including denoising and code fragment representation learning, to better capture structural and semantic patterns.

DeepSeekCoder V2 Lite adopts a GPT-style decoder architecture and is instruction-tuned on a large multilingual code corpus, reflecting recent trends in large-scale generative code models.

Together, these models span encoder and decoder architectures, multiple pretraining strategies, and model scale, enabling us to assess whether the abstractions surfaced by our framework generalize across these criteria.

\subsection{Datasets}
We conducted our concept analysis experiments on two datasets. While we report results using Project CodeNet~\cite{puri2021codenet}, we also experimented with the CodeSyntax dataset~\cite{shen-etal-2022-codesyntax}. However, CodeSyntax introduced additional confounding factors—such as novel API usage and inconsistent naming patterns—which made it harder to perform perturbation studies and see clear trends. More importantly, it lacked the structured metadata available in CodeNet, making it difficult to construct downstream tasks that align cleanly with our fine-tuning setup. As a result, comparisons between pretraining and fine-tuning conditions were less reliable on CodeSyntax.

In contrast, CodeNet’s rich annotations and controlled diversity enabled us to define downstream tasks with consistent format and semantics. This consistency was essential for analyzing the effect of fine-tuning relative to the pretrained model's latent abstractions.

We show some examples and syntactic alignment results from CodeSyntax in the supplementary material. 

\section{Annotation Guidelines}
\label{Annotation Guidelines}

\subsection{Prompt Engineering Impact}
To evaluate the effect of prompt engineering on label quality, we conducted a targeted analysis using a structured questionnaire completed by a single expert annotator familiar with the predefined taxonomy. The goal was to assess whether prompt modifications improved the quality of GPT-4o's generated labels in comparison to prior versions and human annotations.

\vspace{0.5em}
\noindent The annotator responded to the following three questions for each evaluated cluster:

\begin{itemize}
    \item \textbf{Question:} Has prompt engineering made V1 unacceptable label acceptable?
    \item \textbf{Options:} \texttt{["N/A", "Yes", "No"]}
    \item \textbf{Follow-up:} If \texttt{"No"} is selected, please answer: \\
    \textit{Describe the issues:}
\end{itemize}

\begin{itemize}
    \item \textbf{Question:} Is GPT-4o's syntactic label superior to the human label?
    \item \textbf{Options:} \texttt{["Yes", "No", "Same"]}
    \item \textbf{Follow-up:} If \texttt{"No"} is selected, please answer: \\
    \textit{Why is it not superior?}
\end{itemize}

\begin{itemize}
    \item \textbf{Question:} Are GPT-4o's semantic tags superior to human tags? (At least 3 tags should be really good and better)
    \item \textbf{Options:} \texttt{["Yes", "No", "Same"]}
    \item \textbf{Follow-up:} If \texttt{"No"} is selected, please answer: \\
    \textit{Why are they not superior?}
\end{itemize}

\subsection{Latent Concept Attribution Impact}

\subsubsection{Inter-Annotator Agreement Study}
\label{sec:inter_annotator_agreement}

To assess the usefulness of the latent concept attribution methods we 
conducted an inter-annotator agreement (IAA) study. Annotators were presented with a code snippet (a sentence), the top salient token discovered using Integrated Gradients for that snippet, the prediction of the downstream task ( programming language classification), the predicted latent concept cluster tokens' word cloud visualization and 
an LLM explanation of the cluster and whether it helps with explaining the prediction. 

Three annotators participated in the study, all of whom were researchers with experience in 
programming languages and software analysis. For each token-cluster pair, annotators were asked 
to make two key judgments: (1) whether the important token alone can explain the prediction in the programming language classification task, and
(2) whether the latent concept cluster helps with language prediction.

To quantify agreement, we computed Fleiss $\kappa$ for the binary judgments (token indication 
cand cluster helpfulness). For the token language indication question, we achieved a 
Fleiss's $\kappa = 0.577$, indicating good agreement. For cluster context helpfulness, the 
Fleiss's $\kappa = 0.525$, suggesting good agreement between annotators.

The results showed that in $37.0\%$ of cases where tokens alone were deemed insufficient for 
language prediction, and that the latent concept clusters provided helpful additional information validating the usefulness of our study. 

Disagreements primarily occurred in cases where tokens had ambiguous usage patterns across 
multiple programming languages and were resolved through discussion. These findings provide empirical support for the usefulness of 
our latent concept attribution  approach for explaining individual predictions reliably. 

\subsection*{Questions Asked to the Annotators}

We presented annotators with the following questions for each token-cluster pair:

\begin{enumerate}[label=\textbf{Q\arabic*.}]
    \item ``Does the (important) token (obtained using Integrated Gradients) by itself indicate which language the code belongs to?''
    \begin{itemize}
        \item Response options: \{\texttt{Yes}, \texttt{No}\}
    \end{itemize}

    \item ``Does having additional concept cluster information help with the prediction?''
    \begin{itemize}
        \item Response options: \{\texttt{Yes}, \texttt{No}\}
    \end{itemize}

\end{enumerate}

\noindent
For each question, annotators were required to select exactly one option from the provided choices. 

Annotators were compensated in the form of research credits.

\section{Complete Prompt for Latent Concept Annotation}
\subsection{Before Prompt Engineering}
\begin{lstlisting}[caption={Prompt 1: Basic Token Labeling}]
Generate a concise label or theme for the following Java code tokens: {token_summary}.
\end{lstlisting}

\begin{lstlisting}[caption={Prompt 2: Functional Token Labeling}]
Given the following Java code tokens: {token_summary} and the corresponding lines of code that use them: {context_summary}, what functionality or pattern do the tokens represent? Provide a concise label for the cluster of given code tokens.
\end{lstlisting}

\subsection{After prompt Engineering}
\label{Appendix: Full_prompt}
\begin{lstlisting}[caption={Prompt used with GPT-4}]
You are analyzing a cluster of Java tokens and their context sentences. Each cluster has one or more unique tokens. Your task is to identify the role or function these tokens play within the context of the provided sentences. Focus on understanding what the tokens are achieving in the code and their syntactic or semantic significance.

**Guidelines for Analysis:**
1. **Tokens:** Review the provided tokens.
2. **Context Sentences:** Examine the context sentences to understand the usage of the tokens.
3. **Role Identification:** Determine the role the tokens play in the context sentences, including their syntactic and semantic significance.
4. **Concise Label:** Choose a descriptive label that accurately describes the function or role of the tokens in the code. Use specific terminology where applicable (e.g., `Buffer Manipulation`, `Method Invocation`, `Parameter Handling`).
5. **Semantic Tags:** Include 3-5 relevant semantic tags that describe what is being achieved in the context sentences.
6. **Description:** Provide a concise description (1-2 sentences) explaining the role of the tokens in the code.
7. ' ( ' would have label 'Opening Parenthesis' and ')' would have label 'Closing Parenthesis'

### Examples from Previous Clusters:
1. **Tokens:** `returnBuffer, concatBuffer`  
   **Context Sentences:**
   - returnBuffer.append(minParam);
   - returnBuffer.append(FieldMetaData.Decimal.SQ_CLOSE);
   - StringBuffer concatBuffer = new StringBuffer();
   - concatBuffer.append(toAdd);
   
   **Label:** Buffer Manipulation  
   **Semantic Tags:** StringBuilder, StringBuffer, Data Aggregation, String Concatenation  
   **Description:** The tokens represent `StringBuilder` and `StringBuffer` objects used for building strings by appending data elements in sequence.

2. **Tokens:** `.`  
   **Context Sentences:**
   - returnBuffer.append(FieldMetaData.Decimal.SQ_CLOSE);
   - jsonObject.getLong(Form.JSONMapping.FORM_TYPE_ID);
   - date.getTime();
   - fileReader.readLine();
   
   **Label:** Method Invocation Operator  
   **Semantic Tags:** Dot Notation, Method Call, Property Access  
   **Description:** The dot (.) operator is used to call methods or access properties of objects in Java.

Based on the provided tokens and context sentences below, analyze the cluster and provide your response in the following JSON format:

{{
    "Label": "Your concise label here",
    "Semantic_Tags": [
        "Tag1",
        "Tag2",
        "Tag3",
        "Tag4",
        "Tag5"
    ],
    "Description": "Your description here."
}}

Tokens: {', '.join(tokens_list)}
All Context Sentences:
{chr(10).join([f"{i + 1}. {sentence}" for i, sentence in enumerate(context_sentences)])}

Ensure your response is in valid JSON format and includes only the JSON object.
"""

\end{lstlisting}

 \section{Prompt for Local attribution explanation}
\label{Appendix: LACOAT prompt}
\begin{lstlisting}[caption={LLM Prompt for non-CLS Salient Token Explanation}]
The task is Programming Language Classification. The sentence is from {language} code.

Do you find any common semantic, structural, lexical and topical relation between the original token (with its position) given to you and the following list of words? Give a more specific and concise summary about the most prominent relation among these words.

Original token: {highlighted_token}
Token's sentence: {sentence}
Position of the original token in the sentence: {position_idx}
List of words (Cluster): {', '.join(cluster_words)}

Does the List of Words (Cluster) help in predicting that this is {language} code? Why or why not?

Answer to the point

\end{lstlisting}

\begin{lstlisting}[caption={LLM Prompt for CLS Salient Token Explanation}]
[CLS] tokens represent the entire sentence. This sentence is from {language} code. Explain the semantic, structural, lexical, or topical meaning in relation to the list of words from similar contexts. What cohesive meaning does this sentence share with the contextual themes?

Original Sentence: {sentence}
List of cluster words: {', '.join(cluster_words)}

Context Sentences of the list of cluster words:
{context_text}

Answer concisely and to the point about how these patterns are characteristic of {language} code.
\end{lstlisting}

\section{Jaccard Similarity and Cluster Sensitivity Index}
\label{Jaccard}

\subsection{Formal Definition}
Formally, let \( \mathcal{C} = \{c_1, c_2, \ldots, c_K\} \) and \( \mathcal{C}' = \{c_1', c_2', \ldots, c_K'\} \) denote the sets of clusters before and after perturbation. For each pair \( (c_i, c_j') \), we compute the Jaccard similarity as \( \text{Jaccard}(c_i, c_j') = \frac{|c_i \cap c_j'|}{|c_i \cup c_j'|} \), where \( |c_i \cap c_j'| \) is the number of shared tokens and \( |c_i \cup c_j'| \) is the total number of unique tokens across both clusters. To handle permutation invariance, we apply the Hungarian algorithm to find an optimal one-to-one matching \( \pi \) between clusters that maximizes total similarity, i.e., \( \pi = \arg\max_{\pi'} \sum_{i=1}^{K} \text{Jaccard}(c_i, \pi'(c_i')) \), where \( \pi' \) ranges over all bijections between \( \mathcal{C} \) and \( \mathcal{C}' \).

We define the average matched similarity as \( \text{Average Jaccard} = \frac{1}{K} \sum_{i=1}^{K} \text{Jaccard}(c_i, \pi(c_i')) \), and finally compute the Cluster Sensitivity Index as \( \text{CSI} = 1 - \text{Average Jaccard} \). Lower CSI values indicate greater cluster stability and higher robustness to semantic-preserving perturbations.

\subsection{Perturbation Results}

\begin{table}[ht]
\centering
\small
\begin{tabular}{lcc}
\toprule
\textbf{Perturbation} & \textbf{Average Jaccard} & \textbf{CSI} \\
\midrule
\multicolumn{3}{l}{\textbf{Lexical Variations}} \\
\midrule
Deterministic Identifier Renaming & 0.6751 & 0.3249 \\
Identifier Casing Variation & 0.6723 & 0.3277 \\
Minimal Casing Perturbation & 0.6657 & 0.3343 \\
Canonical Identifier Substitution & 1.0000 & 0.0000 \\
\midrule
\multicolumn{3}{l}{\textbf{Structural Modifications}} \\
\midrule
Variable Scope Reassignment & 0.6789 & 0.3211 \\
Instrumentation Insertion & 0.5598 & 0.4402 \\
Boolean Expression Negation & 0.7676 & 0.2324 \\
Pointer Introduction & 0.7342 & 0.2658 \\
Statement Order Randomization & 0.5315 & 0.4685 \\
Switch-to-Conditional Transformation & 0.6986 & 0.3014 \\
No-Op Statement Injection & 0.5845 & 0.4155 \\
\midrule
\multicolumn{3}{l}{\textbf{Fine-Tuned Model Variants}} \\
\midrule
POS-Aware Concept Clustering & 0.3940 & 0.6060 \\
Compile Error-Aware Clustering & 1.0000 & 0.0000 \\
Cross-Language Clustering & 1.0000 & 0.0000 \\
\bottomrule
\end{tabular}
\caption{Average Jaccard similarity and Cluster Sensitivity Index (CSI) for various semantic-preserving perturbations. Lower CSI indicates greater cluster stability.}
\label{tab:perturbation_comparison_grouped}
\end{table}

\newpage

\section{Evolution of concepts across layers}

\begin{figure}[h]
    \centering

    \subcaptionbox[CodeBERT pre-trained]{\scriptsize CodeBERT pre-trained \label{fig:codebert_codenet4k_ast}}{
        \includegraphics[width=0.24\textwidth]{Images/all_plots/codenet_4k.ast/microsoft-codebert-base/alignment_score.pdf}
    }
    \hspace{-0.25cm}
    \subcaptionbox[UnixCoder pre-trained]{\scriptsize UnixCoder pre-trained \label{fig:unixcoder_codenet4k_ast}}{
        \includegraphics[width=0.24\textwidth]{Images/all_plots/codenet_4k.ast/microsoft-unixcoder-base/alignment_score.pdf}
    }
    \hspace{-0.25cm}
    \subcaptionbox[CodeBERT finetuned AST]{\scriptsize CodeBERT finetuned on AST node classification \label{fig:codebert_ast_finetuned}}{
        \includegraphics[width=0.24\textwidth]{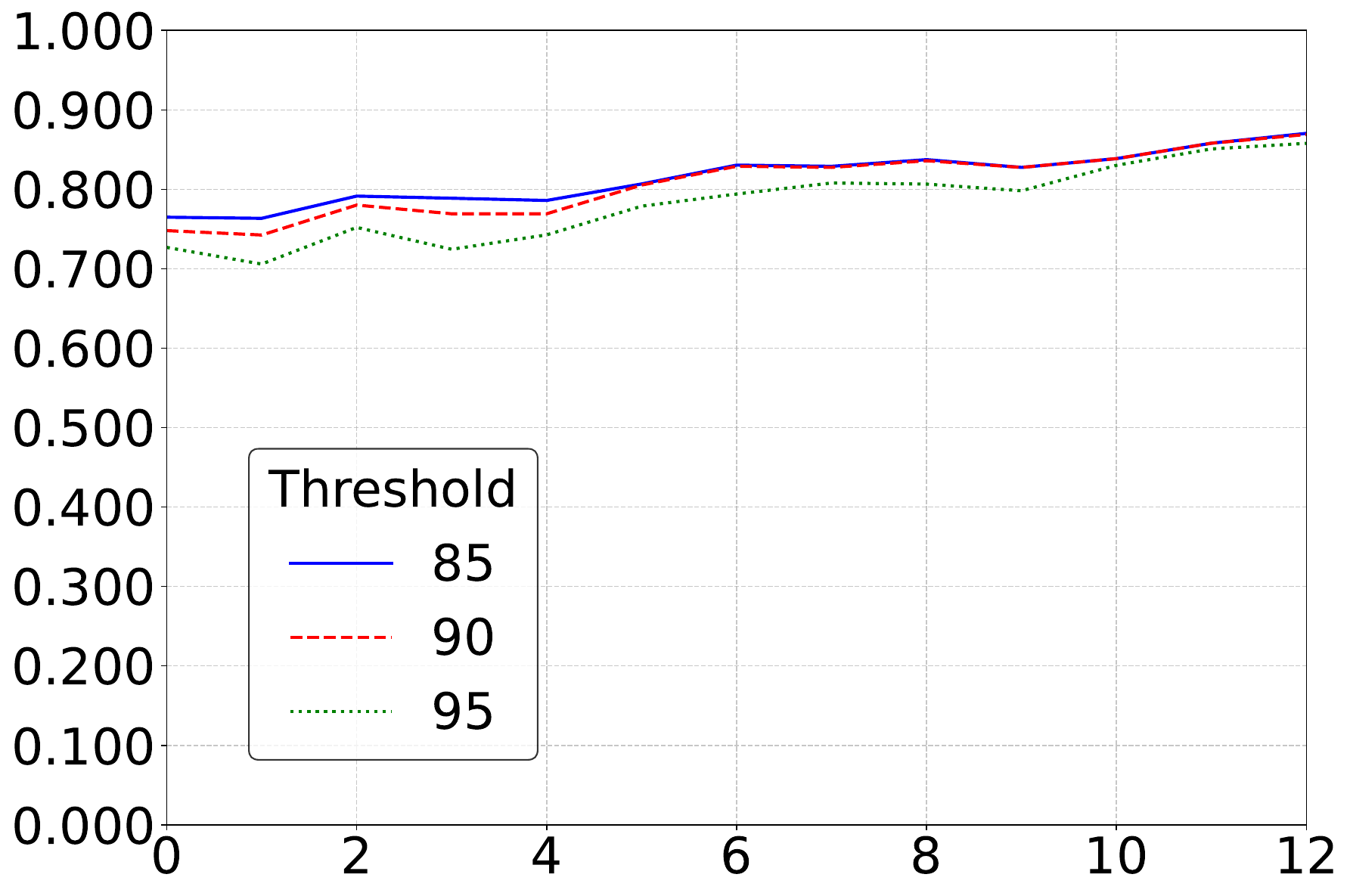}
    }
    \hspace{-0.25cm}
    \subcaptionbox[UnixCoder finetuned AST]{\scriptsize UnixCoder finetuned on AST node classification \label{fig:unixcoder_ast_finetuned}}{
        \includegraphics[width=0.24\textwidth]{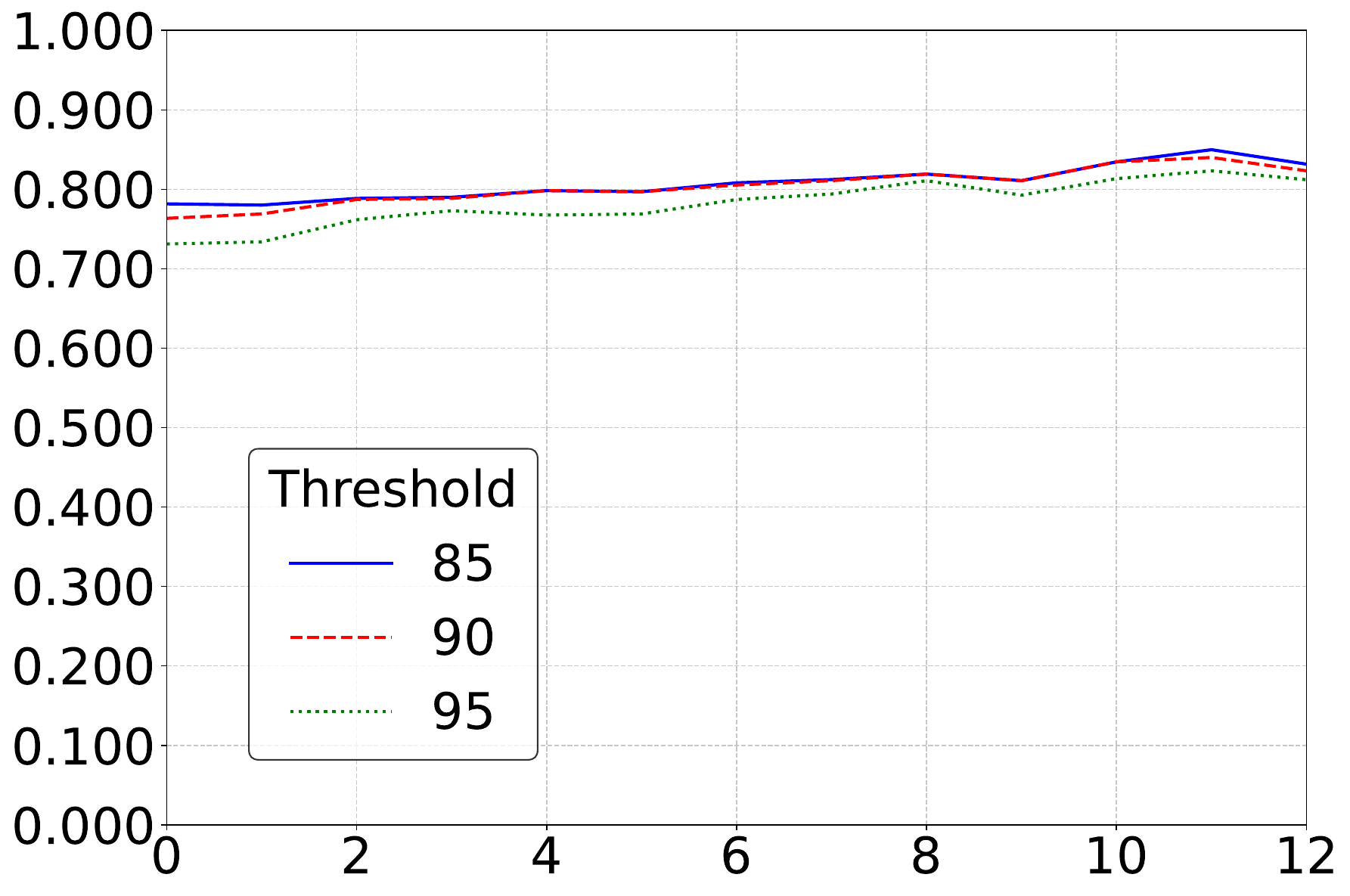}
    }

    \vspace{0.3cm}
    \subcaptionbox[CodeBERT Compile Error]{\scriptsize CodeBERT finetuned on Compile Error Detection \label{fig:codebert_compile_error}}{
        \includegraphics[width=0.24\textwidth]{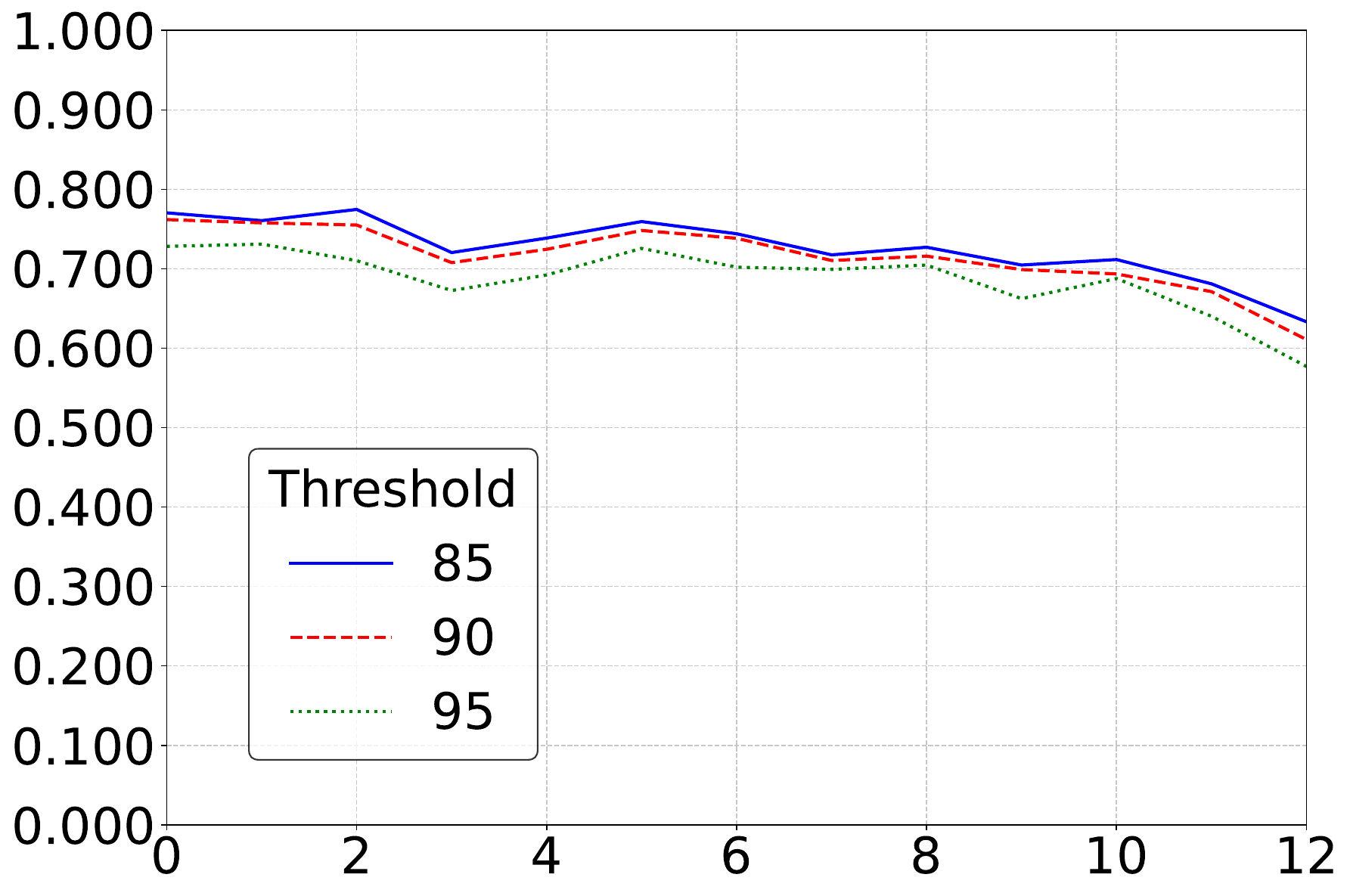}
    }
    \hspace{-0.25cm}
    \subcaptionbox[UnixCoder Compile Error]{\scriptsize UnixCoder finetuned on Compile Error Detection \label{fig:unixcoder_compile_error}}{
        \includegraphics[width=0.24\textwidth]{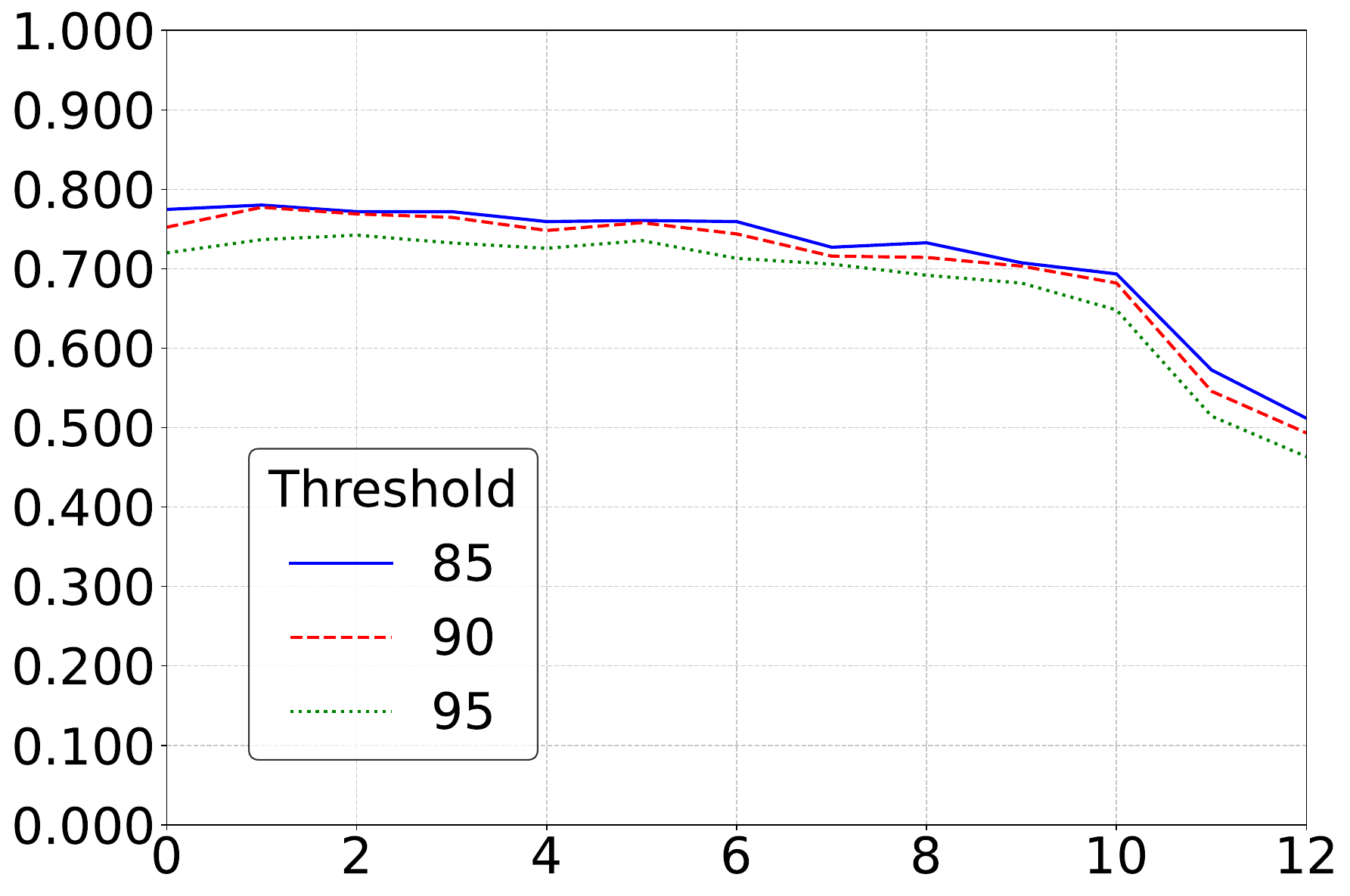}
    }
    \hspace{-0.25cm}
    \subcaptionbox[CodeBERT Lang Cls]{\scriptsize CodeBERT finetuned on Language Classification \label{fig:codebert_lang_cls}}{
        \includegraphics[width=0.24\textwidth]{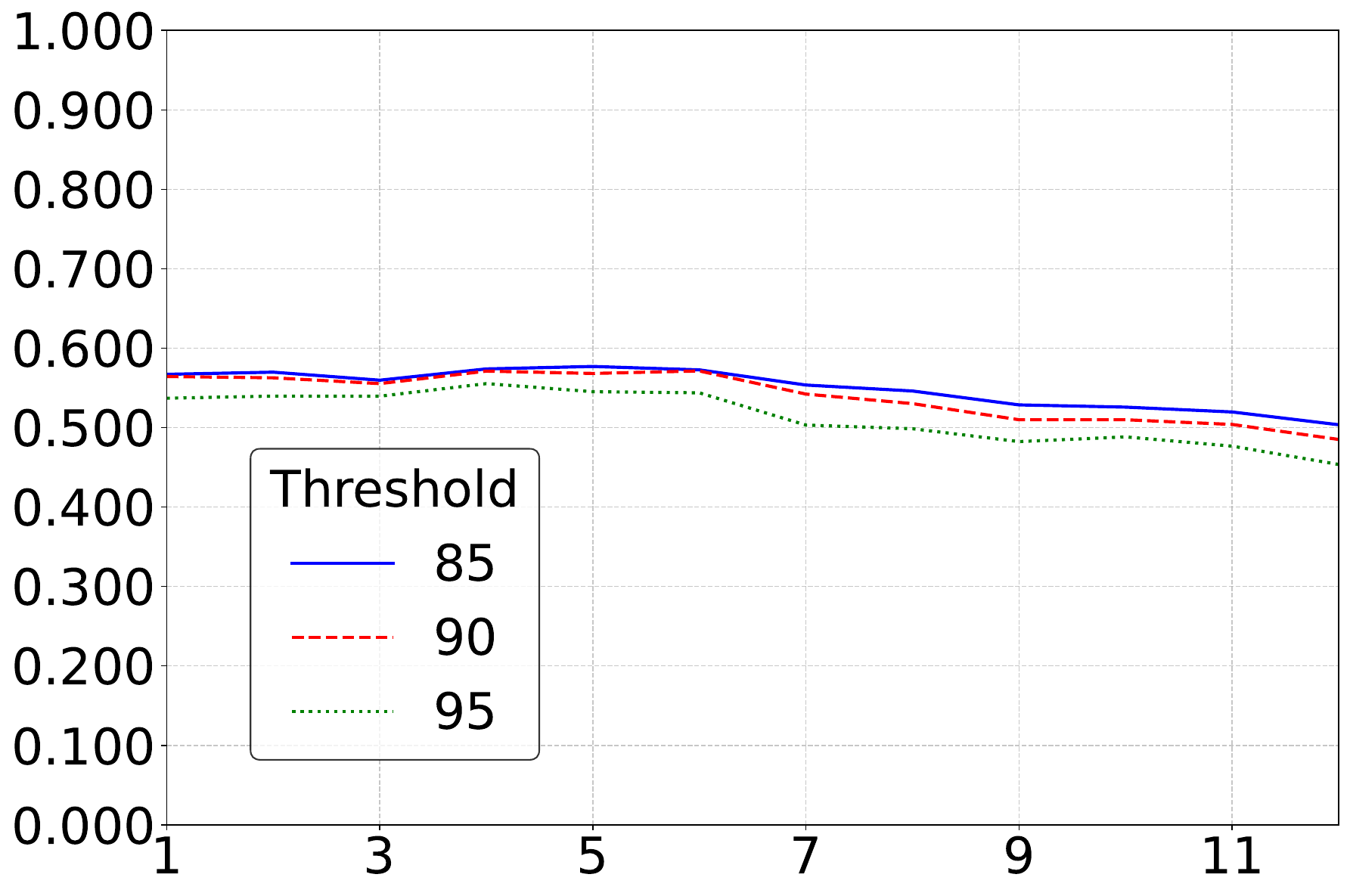}
    }
    \hspace{-0.25cm}
    \subcaptionbox[UnixCoder Lang Cls]{\scriptsize UnixCoder finetuned on Language Classification \label{fig:unixcoder_lang_cls}}{
        \includegraphics[width=0.24\textwidth]{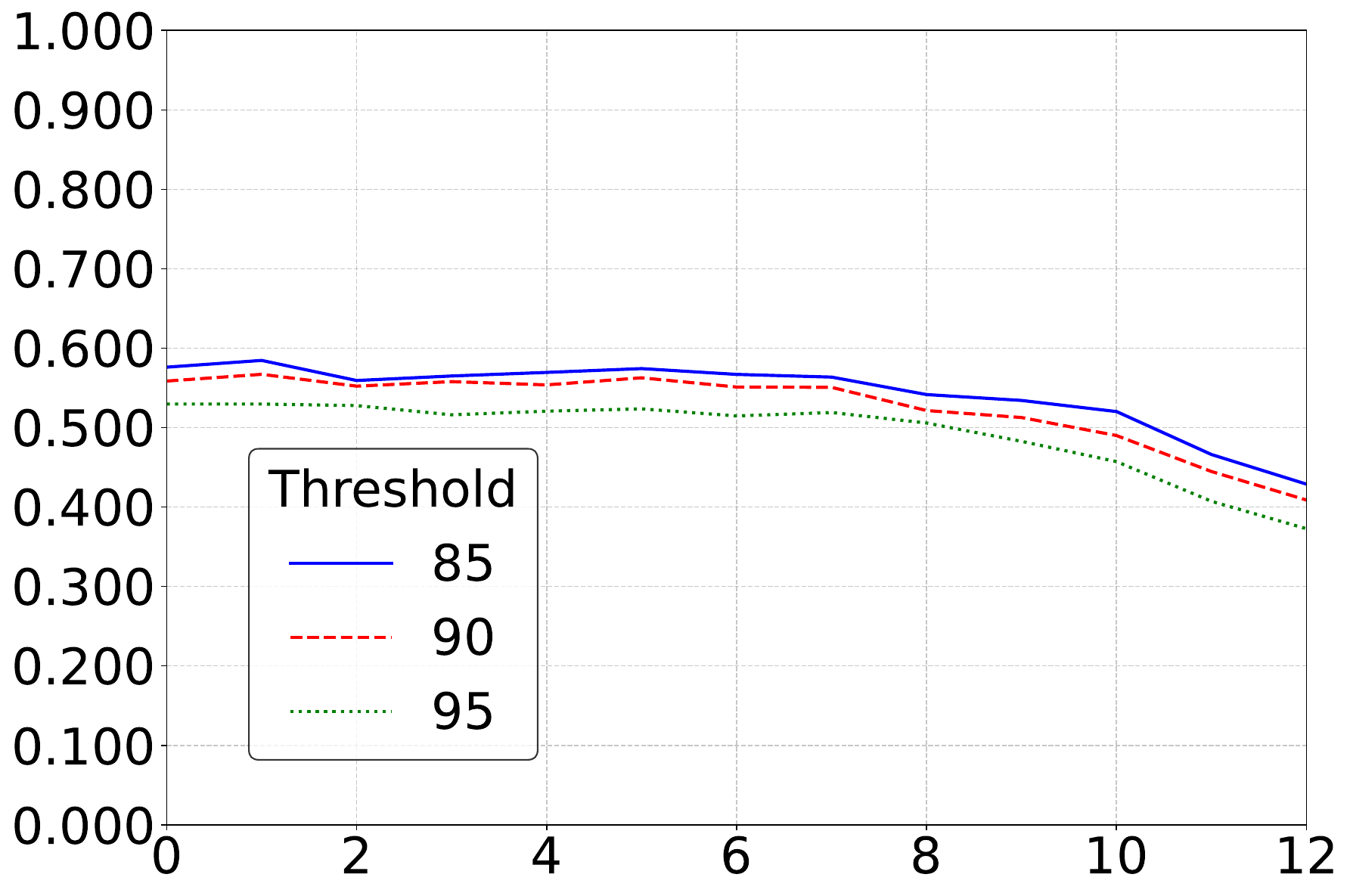}
    }

    \caption{
    Syntactic alignment scores(x axis) across layers(y axis)for CodeBERT and UnixCoder across datasets and fine-tuning objectives. (a–d) show models on CodeNet 4k AST (pretrained and AST-finetuned). (e–f) show Compile Error Detection fine-tuning on CodeNet 4k AST. (g–h) show Language Classification fine-tuning on CodeNet Multi AST. X-axis: layer; Y-axis: alignment score.
    }
    \label{fig:alignment_scores_by_model_and_task}
\end{figure}

\begin{figure}[h]
    \centering

    \subcaptionbox[CodeBERT pre-trained]{\scriptsize CodeBERT pre-trained \label{fig:codebert_codenet4k_ast_lex}}{
        \includegraphics[width=0.24\textwidth]{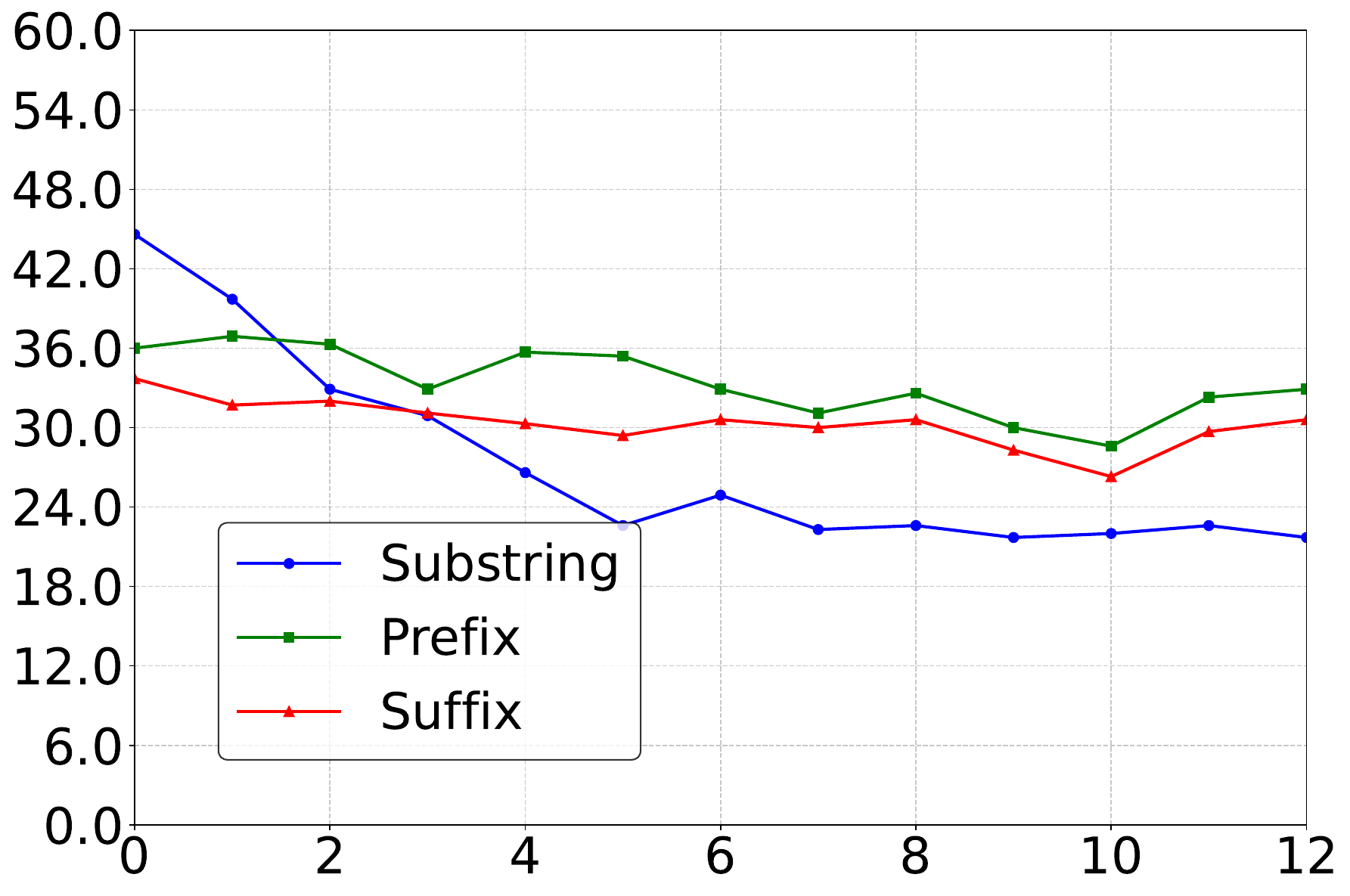}
    }
    \hspace{-0.25cm}
    \subcaptionbox[UnixCoder pre-trained]{\scriptsize UnixCoder pre-trained \label{fig:unixcoder_codenet4k_ast_lex}}{
        \includegraphics[width=0.24\textwidth]{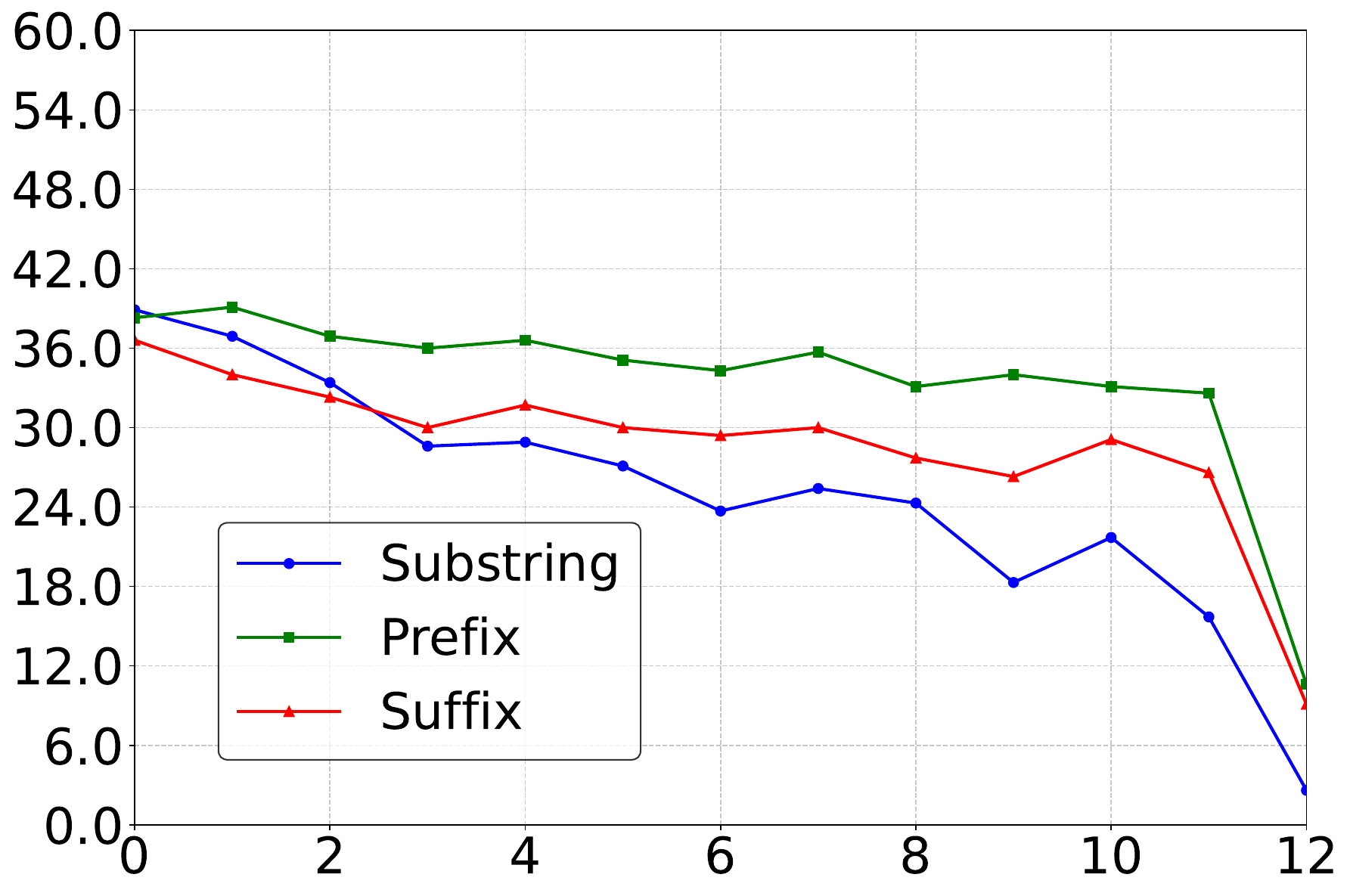}
    }
    \hspace{-0.25cm}
    \subcaptionbox[CodeBERT finetuned AST]{\scriptsize CodeBERT finetuned on AST node classification \label{fig:codebert_ast_finetuned_lex}}{
        \includegraphics[width=0.24\textwidth]{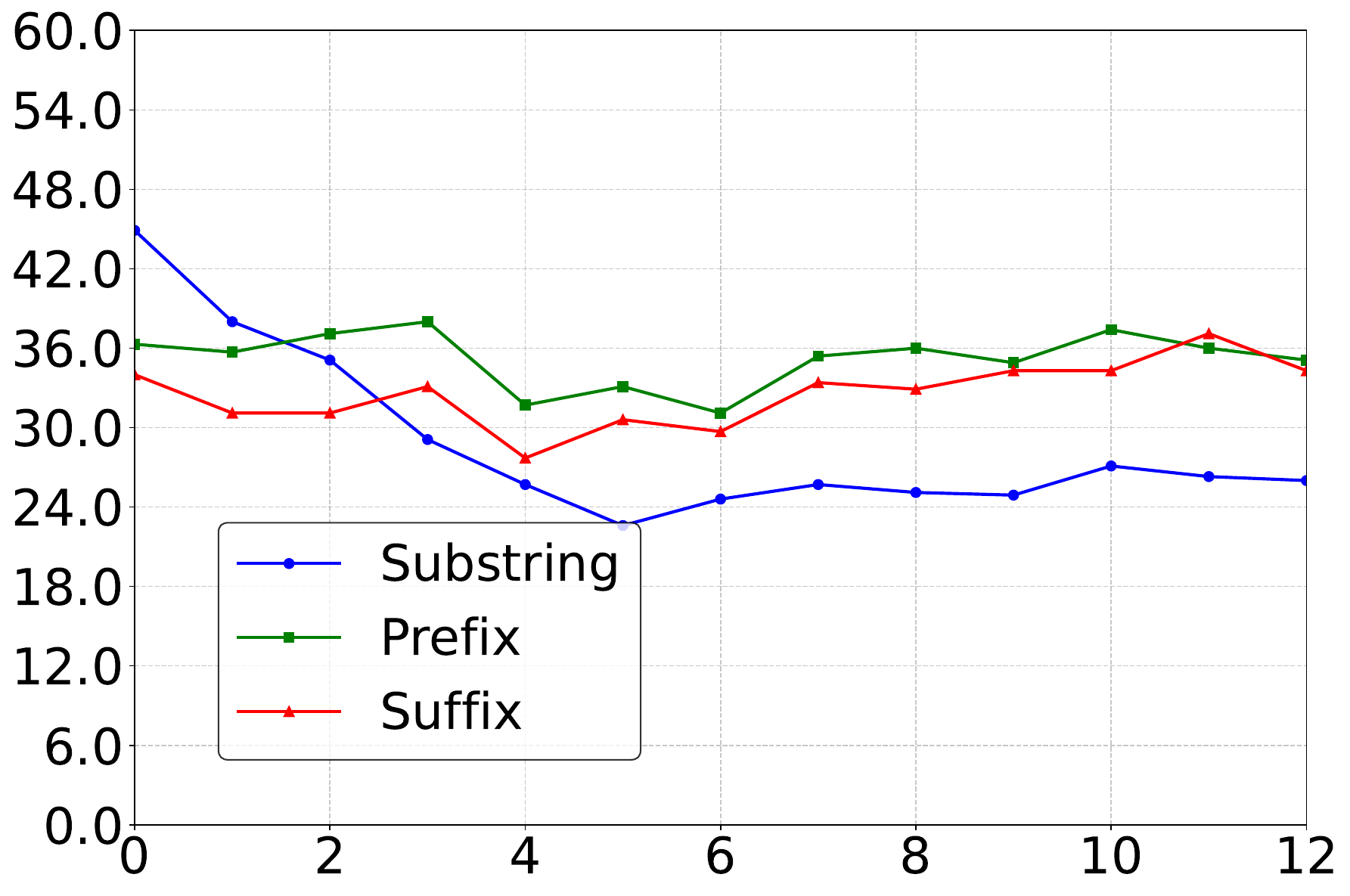}
    }
    \hspace{-0.25cm}
    \subcaptionbox[UnixCoder finetuned AST]{\scriptsize UnixCoder finetuned on AST node classification \label{fig:unixcoder_ast_finetuned_lex}}{
        \includegraphics[width=0.24\textwidth]{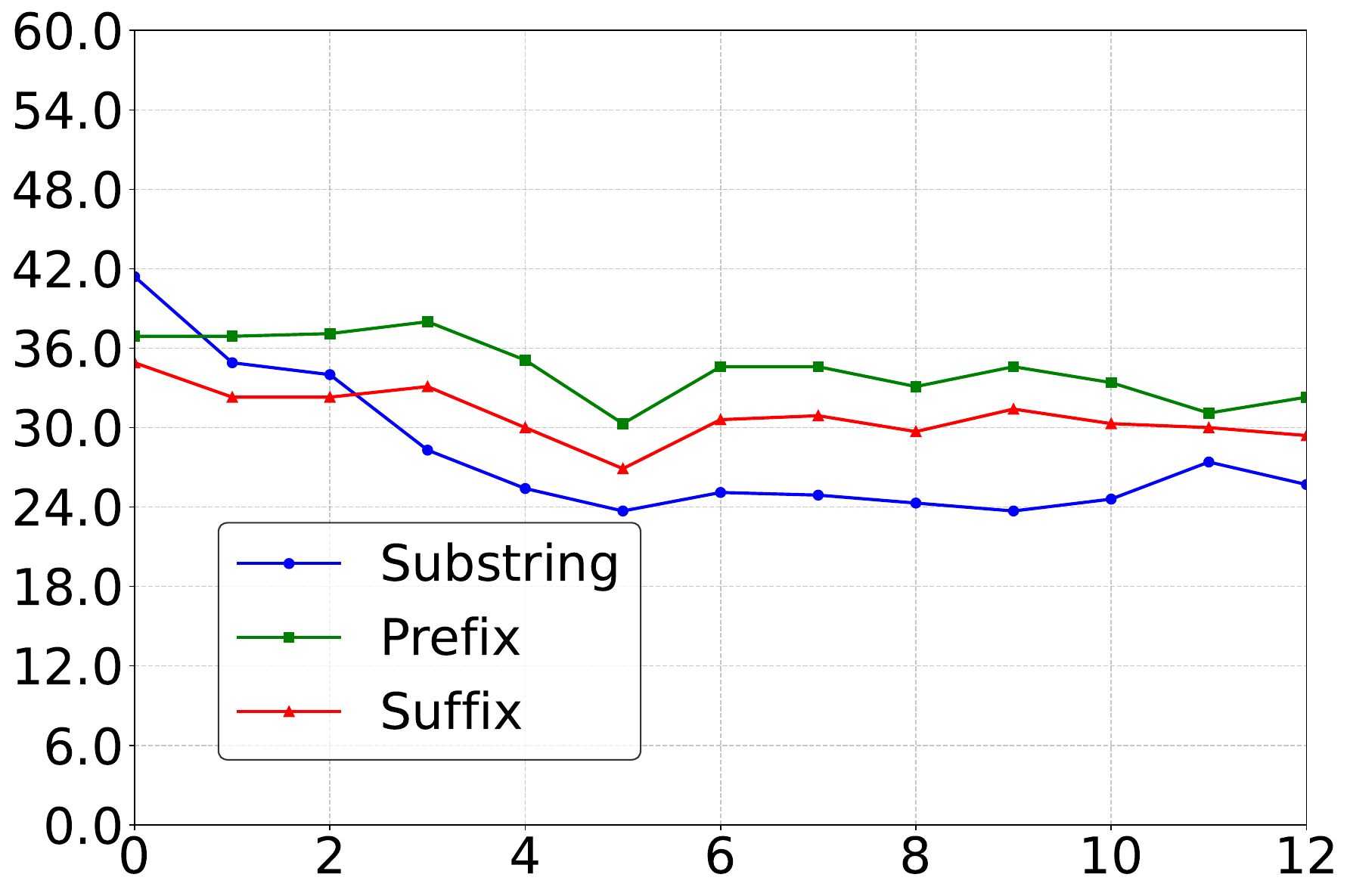}
    }

    \vspace{0.3cm}
    \subcaptionbox[CodeBERT Compile Error]{\scriptsize CodeBERT finetuned on Compile Error Detection \label{fig:codebert_compile_error_lex}}{
        \includegraphics[width=0.24\textwidth]{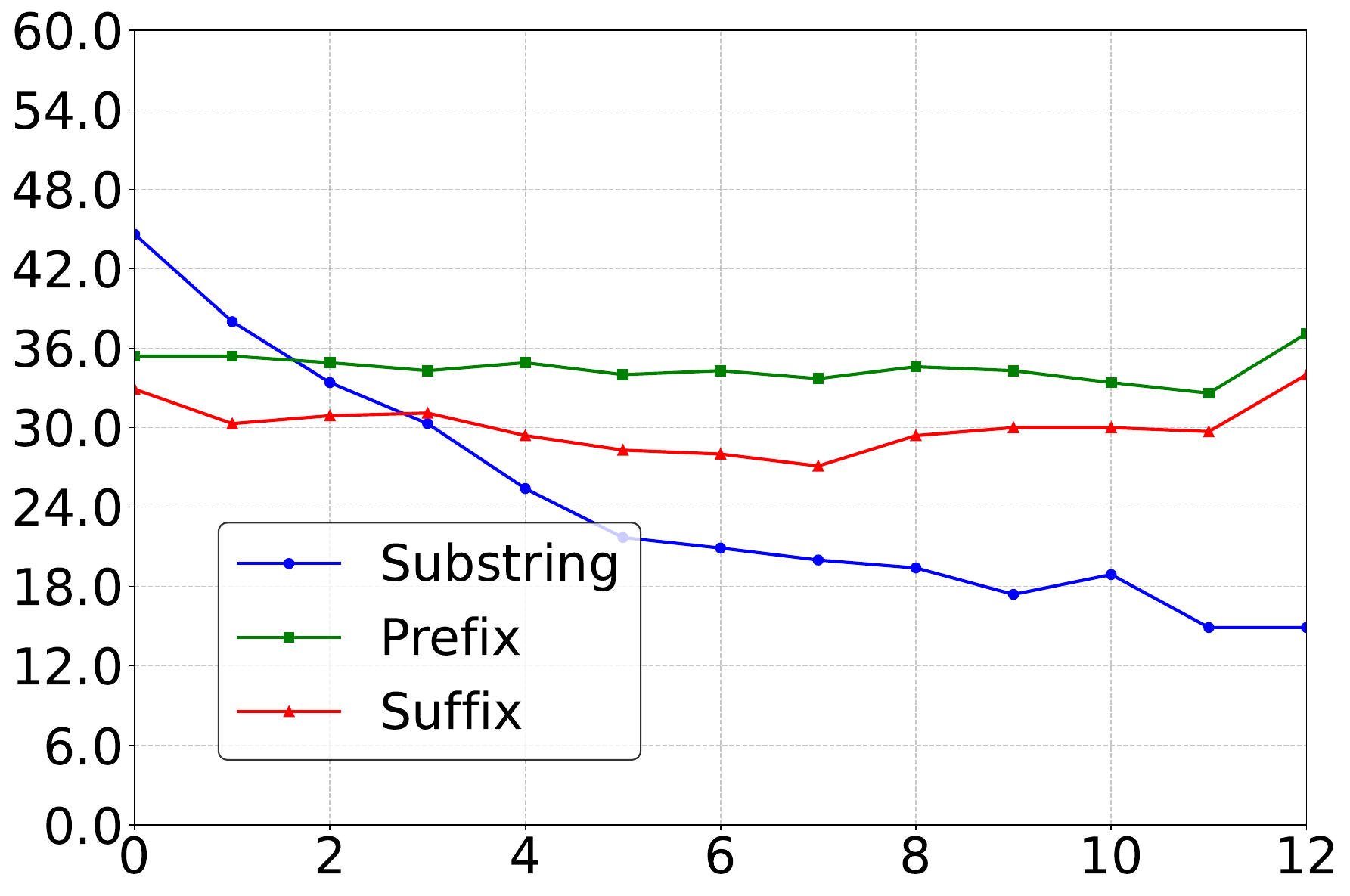}
    }
    \hspace{-0.25cm}
    \subcaptionbox[UnixCoder Compile Error]{\scriptsize UnixCoder finetuned on Compile Error Detection \label{fig:unixcoder_compile_error_lex}}{
        \includegraphics[width=0.24\textwidth]{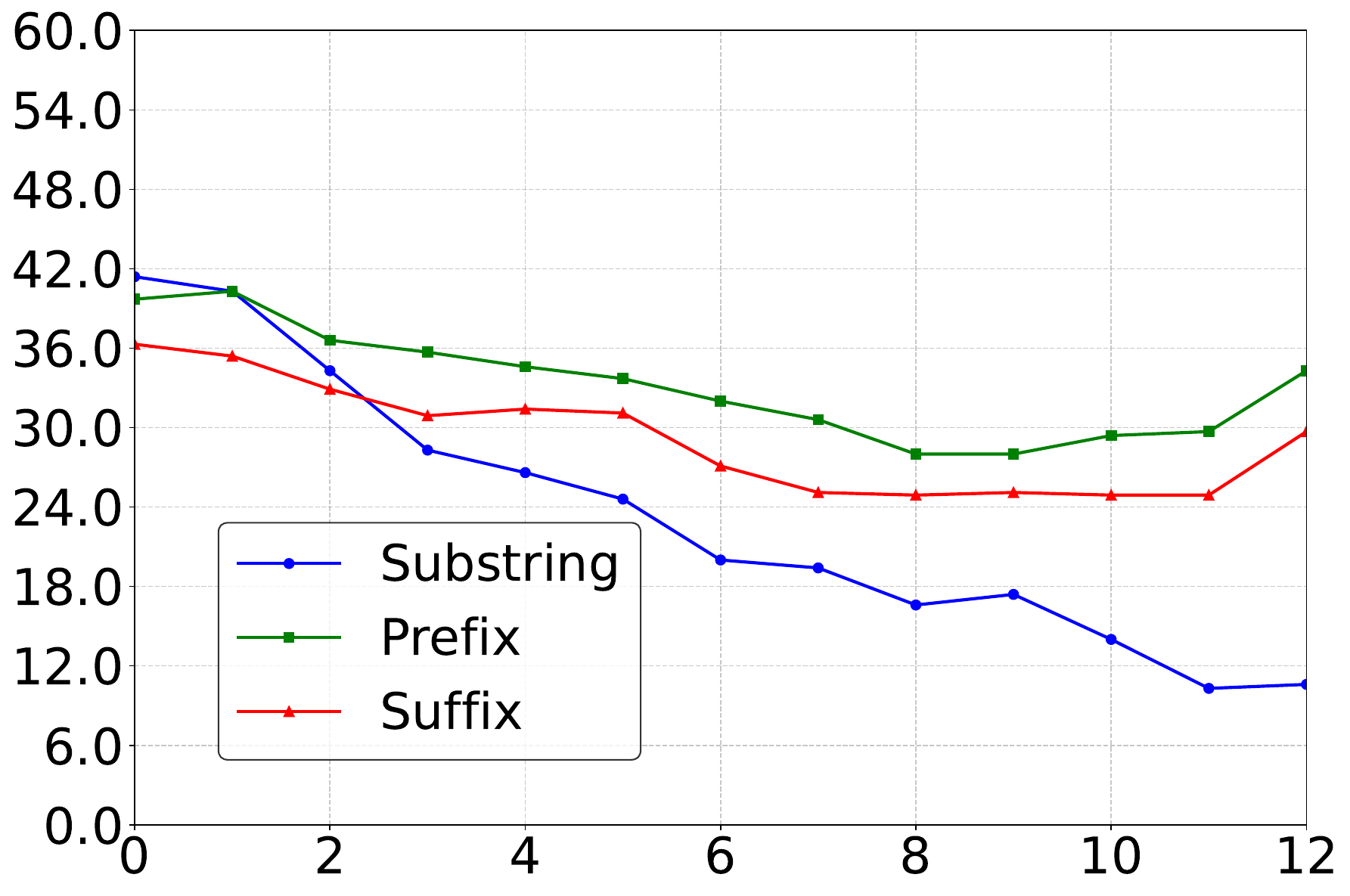}
    }
    \hspace{-0.25cm}
    \subcaptionbox[CodeBERT Lang Cls]{\scriptsize CodeBERT finetuned on Language Classification \label{fig:codebert_lang_cls_lex}}{
        \includegraphics[width=0.24\textwidth]{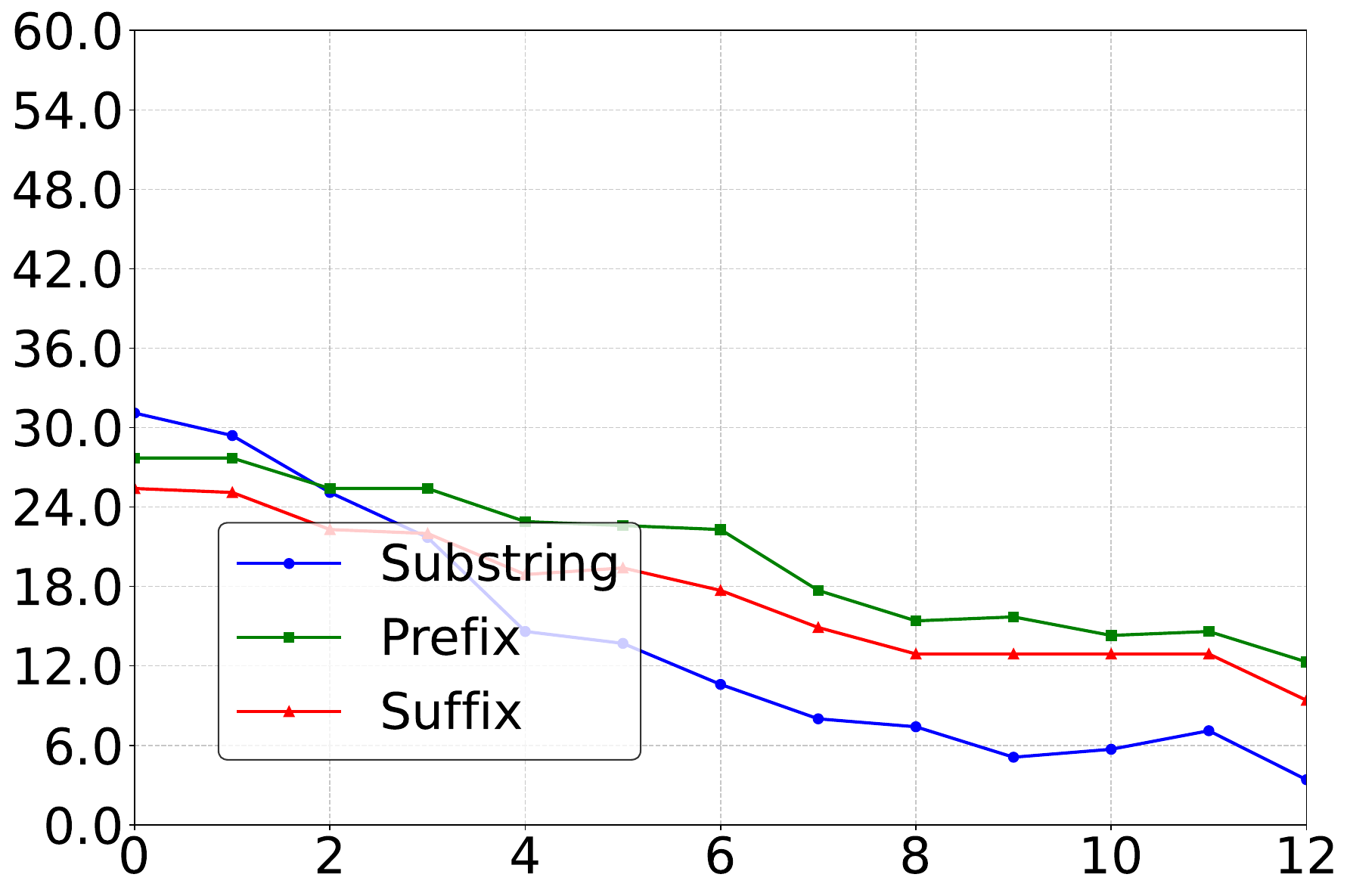}
    }
    \hspace{-0.25cm}
    \subcaptionbox[UnixCoder Lang Cls]{\scriptsize UnixCoder finetuned on Language Classification \label{fig:unixcoder_lang_cls_lex}}{
        \includegraphics[width=0.24\textwidth]{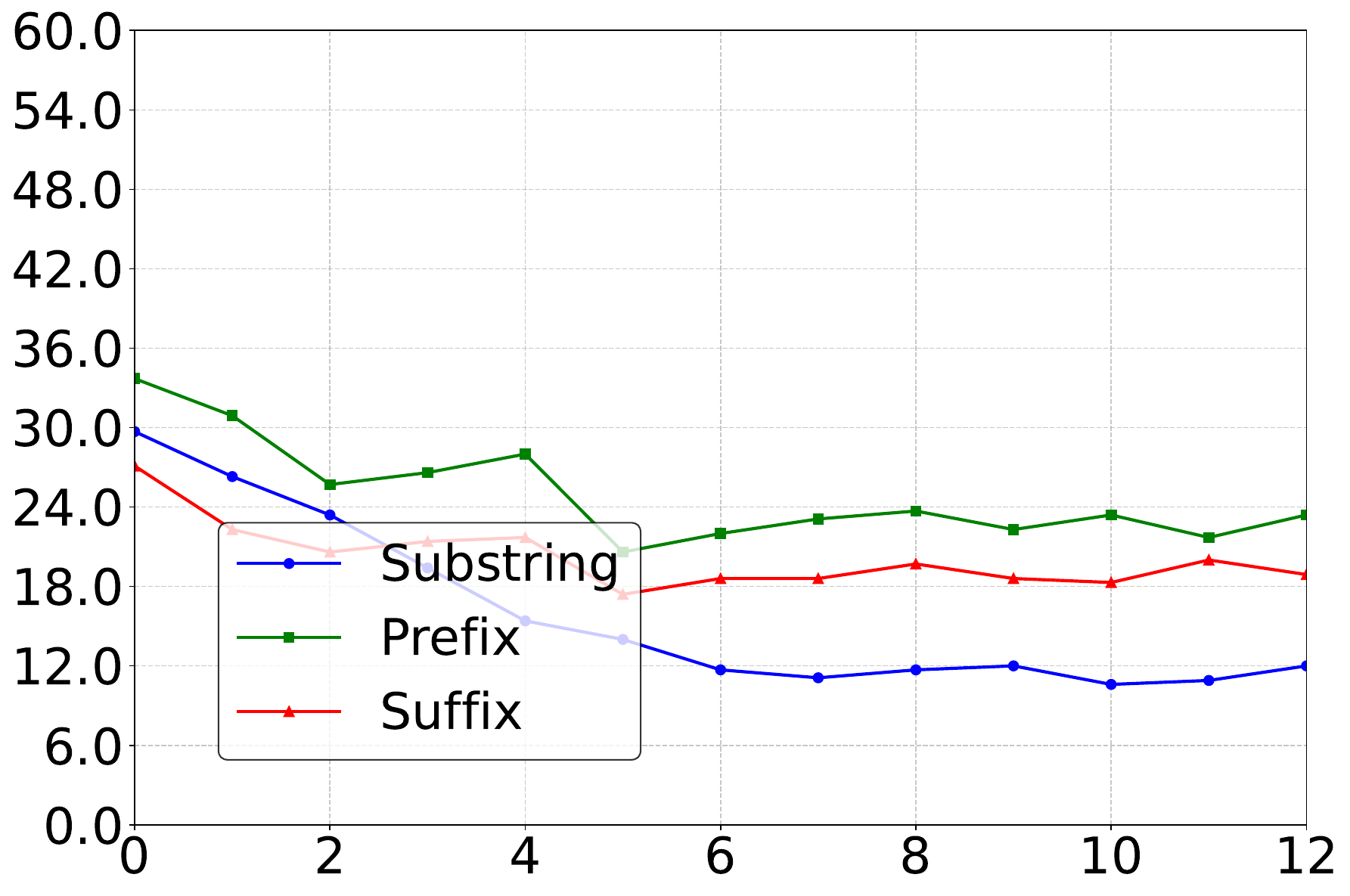}
    }

    \caption{
    Lexical alignment scores for CodeBERT and UnixCoder across datasets and fine-tuning objectives. (a–d) show models on CodeNet 4k AST (pretrained and AST-finetuned). (e–f) show Compile Error Detection fine-tuning on CodeNet 4k AST. (g–h) show Language Classification fine-tuning on CodeNet Multi AST. X-axis: layer; Y-axis: alignment score.
    }
    \label{fig:alignment_scores_by_model_and_task}
\end{figure}

\begin{figure}[htbp]
  \centering
  \begin{subfigure}[b]{0.45\textwidth}
    \centering
    \includegraphics[width=\textwidth]{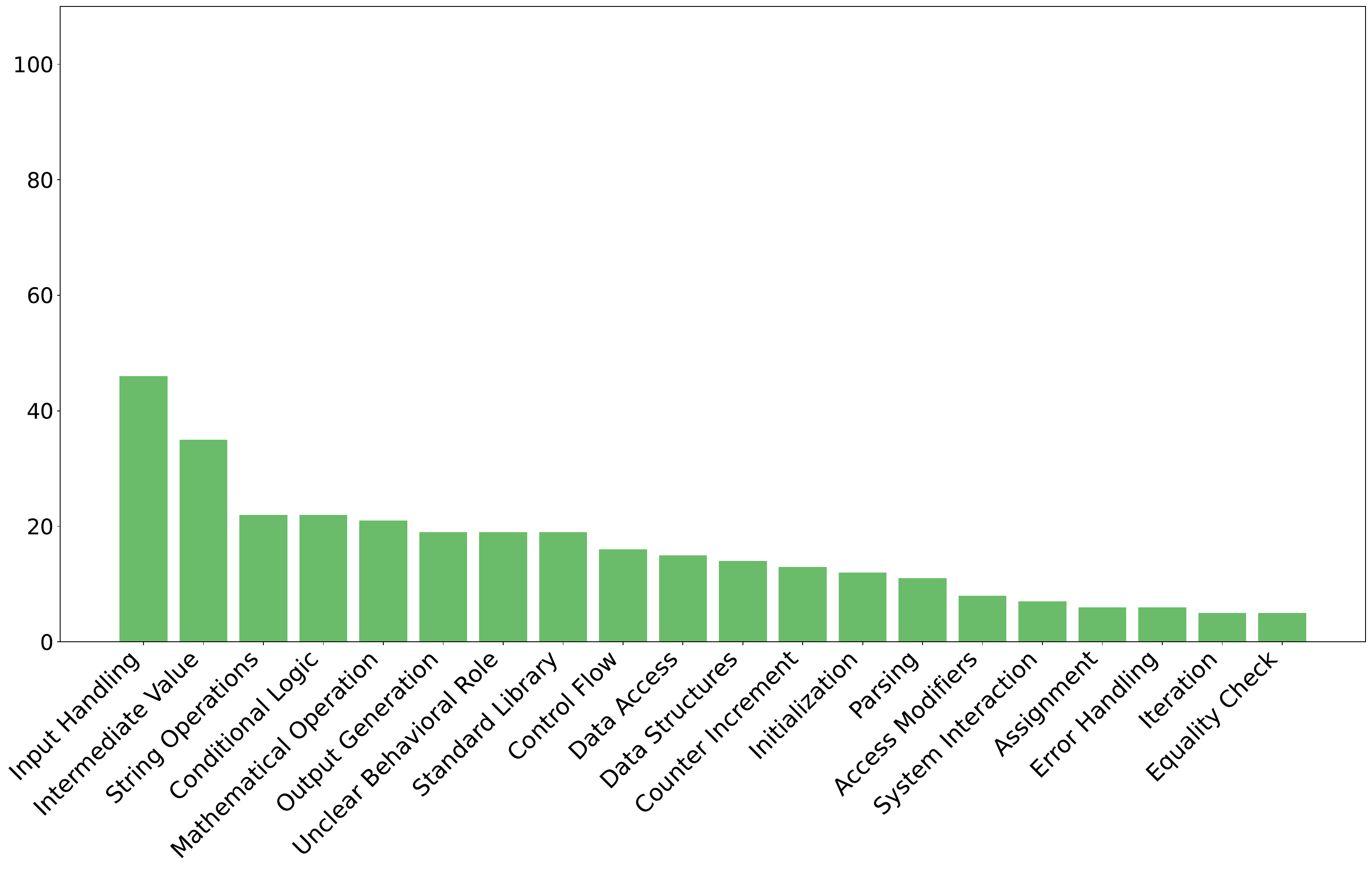}
    \caption{Codebert-base}
  \end{subfigure}
  \hspace{0.05\textwidth}
  \begin{subfigure}[b]{0.45\textwidth}
    \centering
    \includegraphics[width=\textwidth]{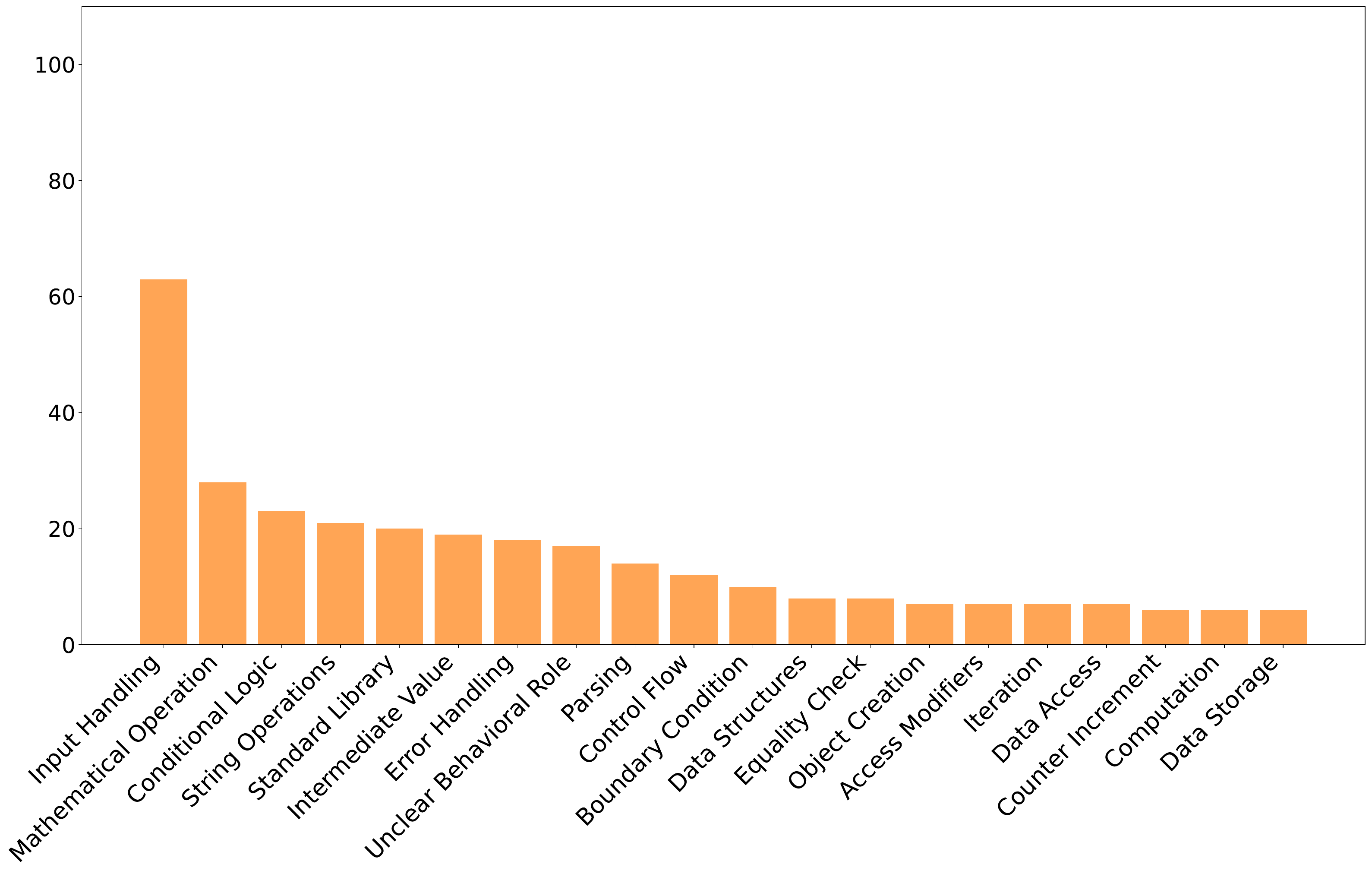}
    \caption{Finetuned on Tree Sitter}
  \end{subfigure}

  \vspace{0.05\textwidth}

  \begin{subfigure}[b]{0.45\textwidth}
    \centering
    \includegraphics[width=\textwidth]{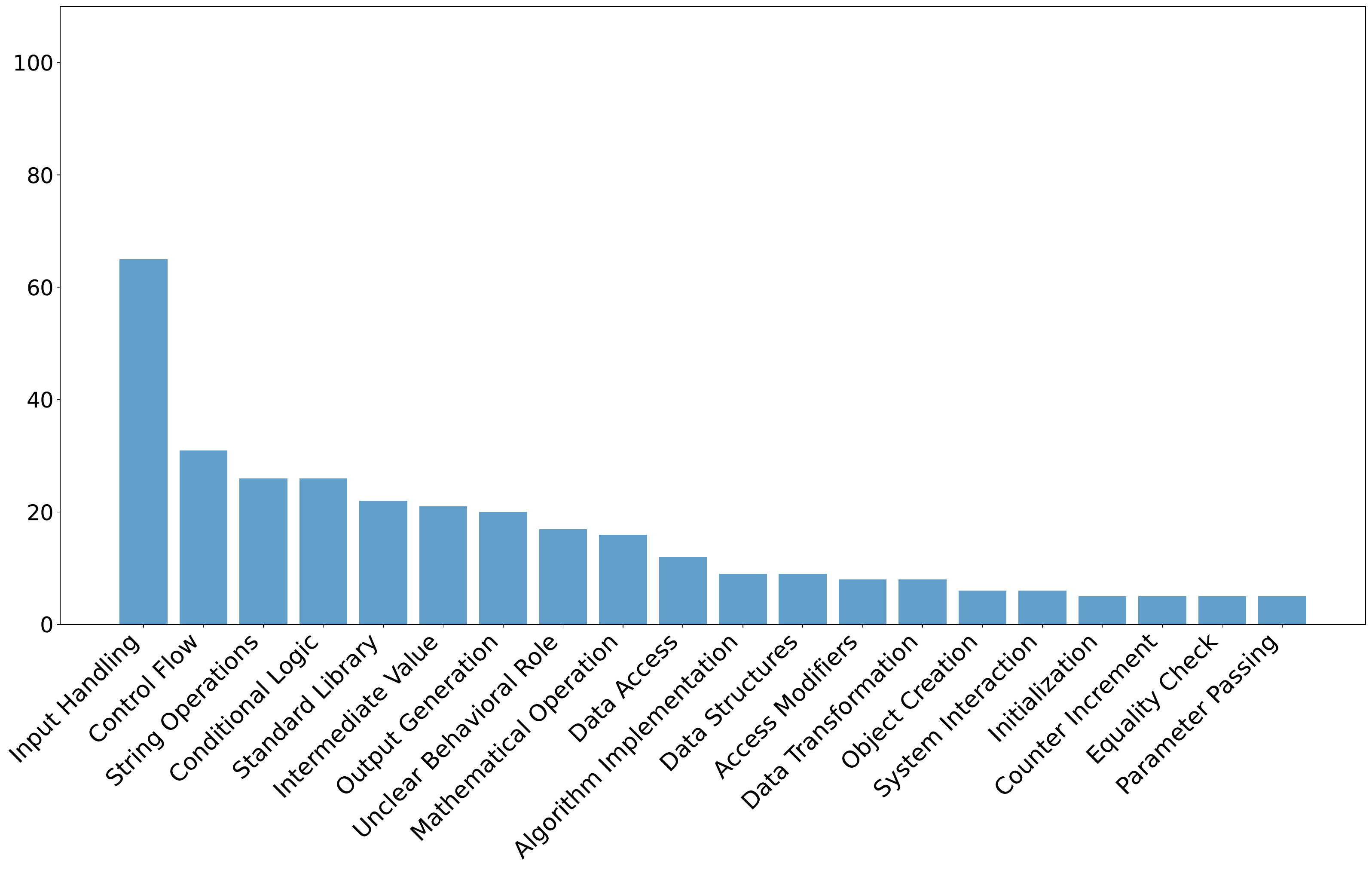}
    \caption{Finetuned on Error Detection}
  \end{subfigure}
  \hspace{0.05\textwidth}
  \begin{subfigure}[b]{0.45\textwidth}
    \centering
    \includegraphics[width=\textwidth]{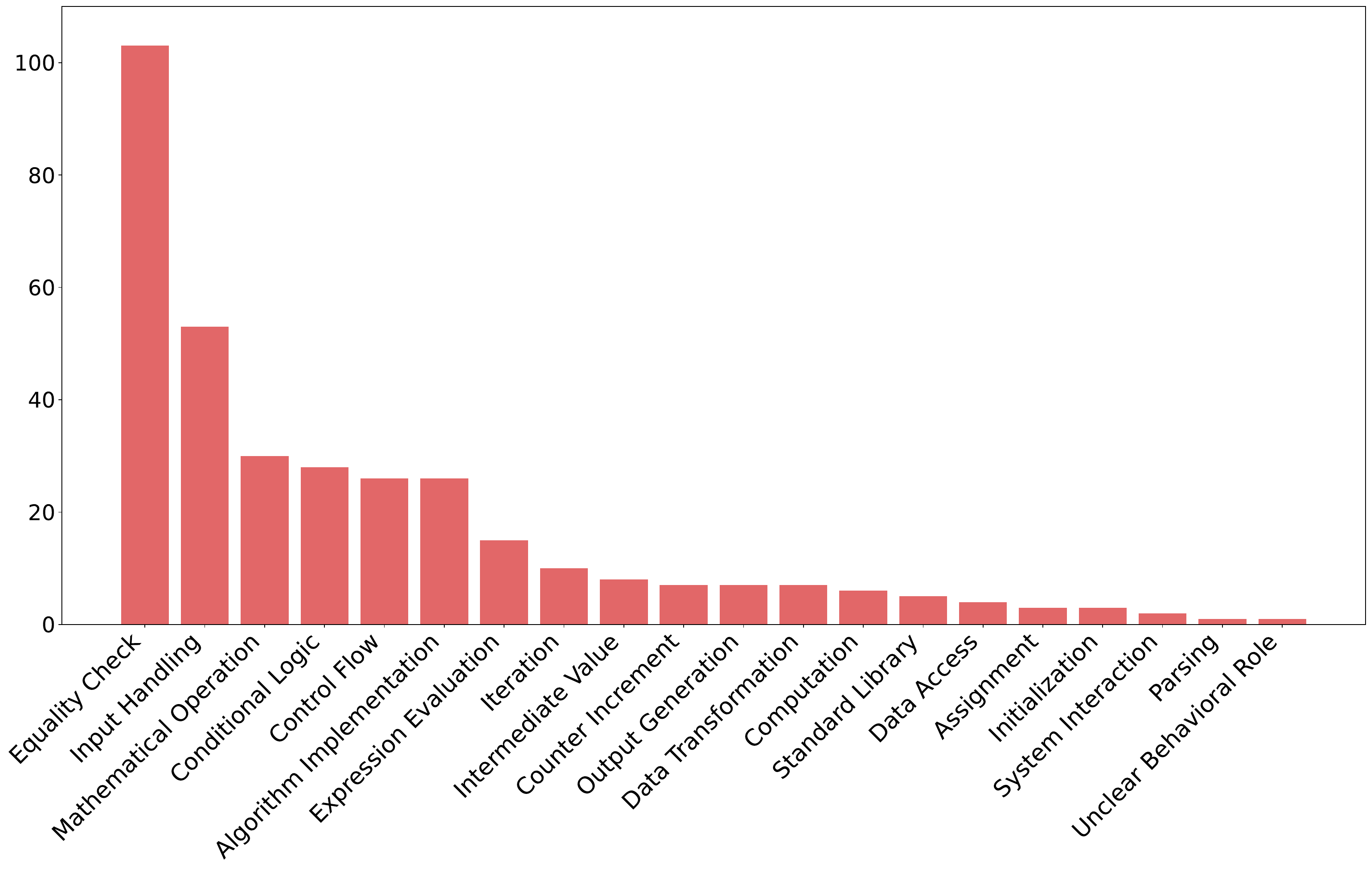}
    \caption{Finetuned on Language Classification}
  \end{subfigure}

  \caption{Top 20 Semantic Tags Across Different CodeBERT Variants}
  \label{fig:two-by-two}
\end{figure}

\newpage
\section{Alignment results for UniXCoder and DeepSeekCoder}
\begin{table*}[ht]
    \centering
    \scriptsize
    \setlength{\tabcolsep}{3pt}
    \begin{tabular}{l|c|c|c|c}
        \toprule
        \rowcolor[gray]{0.9}
        \textbf{Lexical Pattern (\%)} 
        & \textbf{Pre-trained} 
        & \makecell{\textbf{Finetuned on AST} \\ \textbf{Node Classification}} 
        & \makecell{\textbf{Finetuned on Compile} \\ \textbf{Error Detection}} 
        & \makecell{\textbf{Finetuned on} \\ \textbf{Language Classification}} \\
        \midrule
        Substring match(>3)  
        & 2.6 & 25.7 & 10.6 & 12.0 \\
        
        Prefix  
        & 10.6 & 32.3 & 34.3 & 22.6 \\
        
        Suffix 
        & 9.1 & 29.4 & 29.7 & 19.4 \\
        \midrule
        Camel Casing 
        & 0.9 & 7.7 & 3.1 & 0.9 \\
        
        Pascal Casing 
        & 3.4 & 7.1 & 10.0 & 4.9 \\
        \bottomrule
    \end{tabular}
    \caption{UniXCoder- Lexical alignment of token clusters from layer 12 (\textbf{350 clusters}) across pretraining and fine-tuning stages. The first group captures substring, prefix, and suffix similarity; the second focuses on identifier casing patterns (Camel and Pascal). Evaluated at 80\% similarity threshold.}
    \label{tab:lexical_patterns_unixcoder}
\end{table*}

\begin{table*}[ht]
    \centering
    \scriptsize
    \setlength{\tabcolsep}{3pt}
    \begin{tabular}{l|ccc|ccc|ccc|ccc}
        \toprule
        \rowcolor[gray]{0.9}
        \textbf{Metric} 
        & \multicolumn{3}{c|}{\textbf{Pre-trained}} 
        & \multicolumn{3}{c|}{\makecell{\textbf{Finetuned on AST} \\ \textbf{Node Classification}}}
        & \multicolumn{3}{c|}{\makecell{\textbf{Finetuned on Compile} \\ \textbf{Error Detection}}}
        & \multicolumn{3}{c}{\makecell{\textbf{Finetuned on} \\ \textbf{Language Classification}}} \\
        
        \rowcolor[gray]{0.9}
        & 85\% & 90\% & 95\%
        & 85\% & 90\% & 95\%
        & 85\% & 90\% & 95\%
        & 85\% & 90\% & 95\% \\
        \midrule
        Clusters Labeled (/350) 
        & 103 & 96 & 88 
        & 344 & 343 & 340 
        & 261 & 248 & 232 
        & 240 & 227 & 212 \\
        
        Tag Coverage (\%) 
        & 8.33 & 6.94 & 6.94 
        & 68.06 & 66.67 & 65.28 
        & 27.78 & 27.78 & 26.39 
        & 14.72 & 14.72 & 14.72 \\
        
        Overall Alignment Score 
        & 0.1888 & 0.1719 & 0.1604 
        & 0.8317 & 0.8233 & 0.8121
        & 0.5117 & 0.4932 & 0.4634 
        & 0.4165 & 0.3979 & 0.3765 \\
        
        Unique Tags Identified 
        & 6 & 5 & 5 
        & 49 & 48 & 47 
        & 20 & 20 & 19 
        & 24 & 24 & 24 \\
        
        None Labels 
        & 247 & 254 & 262 
        & 6 & 7 & 10 
        & 89 & 102 & 118 
        & 110 & 123 & 24 \\
        \bottomrule
    \end{tabular}
    \caption{UniXCoder: Comparison of cluster metrics for: pretraining and three types of finetuning, evaluated at varying confidence thresholds.}
    \label{tab:alignment_metrics_codenet_treesitter_unixcoder}
\end{table*}

\begin{table*}[ht]
    \centering
    \scriptsize
    \setlength{\tabcolsep}{4pt}
    \begin{tabular}{l|c}
        \toprule
        \rowcolor[gray]{0.9}
        \textbf{Lexical Pattern (\%)} 
        & \textbf{DeepSeek-Coder-V2-Lite-Instruct (Pre-trained)} \\
        \midrule
        Substring match (>3)  
        & 8.0 \\
        
        Prefix  
        & 26.0 \\
        
        Suffix 
        & 20.6 \\
        \midrule
        Camel Casing 
        & 1.7 \\
        
        Pascal Casing 
        & 8.6 \\
        \bottomrule
    \end{tabular}
    \caption{Lexical alignment of token clusters from layer 12 (\textbf{350 clusters}) of \textbf{DeepSeek-Coder-V2-Lite-Instruct}, evaluated at 80\% similarity threshold.}
    \label{tab:lexical_alignment_deepseek}
\end{table*}

\begin{table*}[ht]
    \centering
    \scriptsize
    \setlength{\tabcolsep}{4pt}
    \begin{tabular}{l|c|c|c}
        \toprule
        \rowcolor[gray]{0.9}
        \textbf{Syntactic Metric} 
        & \multicolumn{3}{c}{\textbf{DeepSeek-Coder-V2-Lite-Instruct (Pre-trained)}} \\
        \rowcolor[gray]{0.9}
        & \textbf{85\%} 
        & \textbf{90\%} 
        & \textbf{95\%} \\
        \midrule
        Clusters Labeled (/350) 
        & 285 & 274 & 250 \\
        
        Tag Coverage (\%) 
        & 29.17 & 27.78 & 26.39 \\
        
        Overall Alignment Score 
        & 0.5530 & 0.5303 & 0.4891 \\
        
        Unique Tags Identified 
        & 21 & 20 & 19 \\
        
        None Labels 
        & 65 & 76 & 100 \\
        \bottomrule
    \end{tabular}
    \caption{Syntactic alignment of token clusters from layer 27 (\textbf{350 clusters}) of \textbf{DeepSeek-Coder-V2-Lite-Instruct}, evaluated at varying confidence thresholds.}
    \label{tab:syntactic_alignment_deepseek}
\end{table*}

\newpage

\section{Dataset Details}
\label{Dataset details}

\begin{table}[h]
  \caption{Dataset sizes (lines of code) for each fine-tuning task. All splits are balanced and derived from Project CodeNet across 200–500 problems. POS tagging is based on token-level AST node type labels.}
  \label{tab:dataset_splits}
  \centering
  \small
  \begin{tabular}{p{4.9cm}ccc}
    \toprule
    \textbf{Task} & \textbf{Train} & \textbf{Validation} & \textbf{Test} \\
    \midrule
    CodeBERT – POS Tagging (Tree-sitter) & 3600 & 400 & -- \\
    CodeBERT – Language Classification    & 3380 & 840 & -- \\
    CodeBERT – Compile Error Detection    & 3200 & 800 & -- \\
    \bottomrule
  \end{tabular}
\end{table}

\begin{table}[h]
  \caption{Dataset sizes (lines of code) for local concept attribution, derived from Project CodeNet. Each row represents a set of code submissions per task.}
  \label{tab:local_explanation_splits}
  \centering
  \small
  \begin{tabular}{p{5.2cm}ccc}
    \toprule
    \textbf{Explanation Task} & \textbf{Train} & \textbf{Validation} & \textbf{Test} \\
    \midrule
    Compile Error Explanation (Saliency)       & 3600 & 400 & 1000 \\
    Language Classification Explanation        & 3780 & 420 & 1200 \\
    \bottomrule
  \end{tabular}
\end{table}

\begin{table}[h]
  \caption{Activation extraction datasets used for clustering, syntactic alignment, and lexical alignment tasks. Dataset sizes are measured in lines of code.}
  \label{tab:activation_extraction_sizes}
  \centering
  \small
  \begin{tabular}{p{5.5cm}c}
    \toprule
    \textbf{Activation Extraction Dataset} & \textbf{Size (LOC)} \\
    \midrule
    Fine-Grained Tree-sitter (Token-level) & 4000 \\
    Coarse-Grained Javalang                & 4000 \\
    Multi-Language Tree-sitter             & 4200 \\
    \bottomrule
  \end{tabular}
\end{table}

\newpage
\section{Compute resources}
\label{Compute resources}
All experiments were conducted on an internal high-performance computing (HPC) cluster using nodes equipped with 256 GB RAM and NVIDIA A100 GPUs on a dataset of approximately 4k code snippets obtained from CodeNet. Total storage for all experimetns is about 3.5TB.

Latent Concept Analysis (LCA), including activation extraction and clustering, was run across three models—CodeBERT (4 experiments, 6 hours each), UniXCoder (4 experiments, 6 hours each), and DeepSeekCoder (1 experiment, 44 hours)—requiring approximately 92 GPU hours and Fine-tuning CodeBERT and UniXCoder on three downstream tasks required 1.5 GPU hours.

The Latent concept attribution pipeline required 1 GPU hour for activation extraction and 4.5 CPU hours for training the  logistic regression classifier,clustering and alignment results. 


In total, our reported experiments used approximately 100 GPU hours and 7.2 CPU hours. Preliminary and exploratory runs are excluded from this total and were significantly larger. They included trials for the value of K, different datasets, fine-tuning tasks, etc. Each experiment would take about 6 hours.

\end{document}